\documentclass[aps,prb,twocolumn,superscriptaddress]{revtex4-2}

\usepackage[colorlinks=true,citecolor=blue]{hyperref} 
\usepackage{graphicx}
\usepackage{bm}
\usepackage{amsmath}
\usepackage{amssymb}

\DeclareGraphicsExtensions{.png,.jpg,.eps}
\usepackage{xcolor}

\begin{document}

\author{Valerii Kachin}
\affiliation{Department of Applied Physics, Aalto University, 02150 Espoo, Finland}
\affiliation{Faculty of Mathematics, Informatics and Mechanics, University of Warsaw, Banacha 2, 02-097, Poland}

\author{Teemu Ojanen}
\affiliation{Computational Physics Laboratory, Physics Unit, Faculty of Engineering and Natural Sciences, Tampere University, FI-33014 Tampere, Finland}
\affiliation{Helsinki Institute of Physics P.O. Box 64, FI-00014, Finland}

\author{Jose L. Lado}
\affiliation{Department of Applied Physics, Aalto University, 02150 Espoo, Finland}

\author{Timo Hyart}
\affiliation{Department of Applied Physics, Aalto University, 02150 Espoo, Finland}
\affiliation{Computational Physics Laboratory, Physics Unit, Faculty of Engineering and Natural Sciences, Tampere University, FI-33014 Tampere, Finland}

\title{Effects of  electron-electron interactions in the Yu-Shiba-Rusinov lattice model}

\begin{abstract}
In two-dimensional superconductors,  Yu-Shiba-Rusinov bound states, induced by the magnetic impurities, extend over long distances giving rise to a long-range hopping model supporting a large number of topological phases with distinct Chern numbers. Here, we study how the electron-electron interactions affect on a mean-field level the selection of the realized Chern numbers and the  magnitudes of the topological energy gaps in this model. We find that in the case of an individual choice of the model parameters the interactions can enhance or reduce the topological gap as well as cause topological phase transitions because of the complex interplay of superconductivity, magnetism, and large spatial extent of the Yu-Shiba-Rusinov states.  By sampling a large number of realizations of Yu-Shiba-Rusinov lattice models with different model parameters, we show  that statistically the interactions have no effect on the realized Chern numbers and typical magnitudes of the topological gaps. 
However, the interactions substantially increase the likelihood of the
largest topological gaps in the tails of the energy gap distribution in comparison to the non-interacting case.
\end{abstract}


\maketitle

\section{Introduction}

\begin{figure}[t!]
    \centering
    \includegraphics[width=0.92\columnwidth]{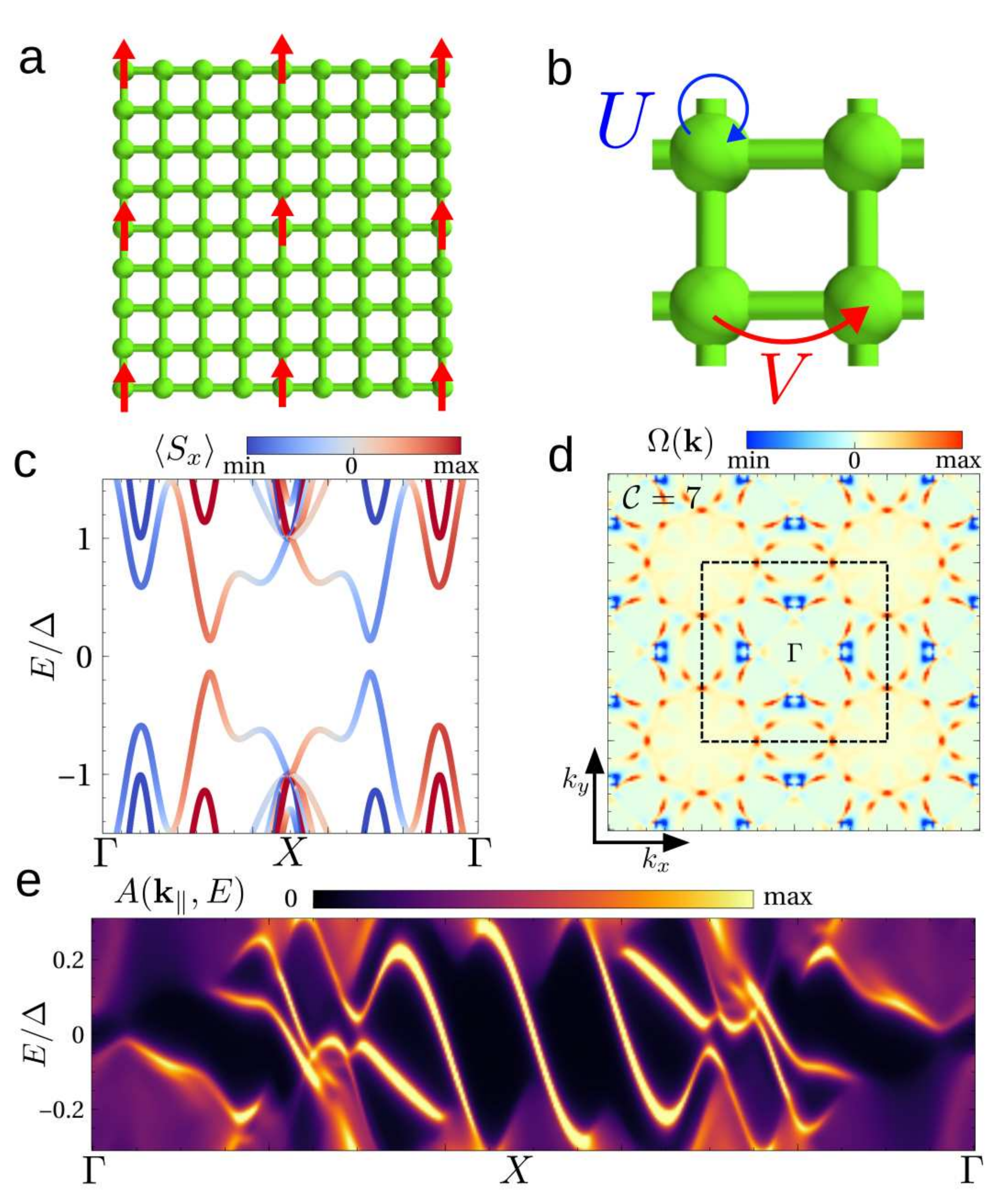}
	\caption{(a) Schematic representation of the Yu-Shiba-Rusinov lattice   consisting of a square lattice of magnetic impurities with a lattice constant $a=4$ placed on the top of a two-dimensional superconductor. 
	(b) The electron-electron interaction terms considered in this work are the onsite $U$ and the nearest-neighbor interaction $V$. (c) Bulk band structure and (d) Berry curvature $\Omega(\mathbf{k})$
	for a Yu-Shiba-Rusinov lattice  with $a=8$, 
	$\nu=0.1$, $J=2.272 t$, $\alpha=0.113 t$, $\Delta=0.06 t$, $U=-0.37 t$, realizing a topological phase with Chern number $C=7$. (e) The surface density of states for the same model parameters  featuring $7$ chiral Majorana edge modes.
}
\label{fig:fig0}
\end{figure}

Magnetic impurities in conventional and non-magnetic impurities in unconventional  superconductors give rise to Yu-Shiba-Rusinov bound states, which have been studied extensively both theoretically and experimentally \cite{yu1965, shiba1968, rusinov1969, BalatskyRMP06, menard2015,grapheneYSR2021}. An important property of these states in two-dimensional systems is that they spatially extent over long distances \cite{menard2015,grapheneYSR2021,KeziYSR2018,PhysRevLett.125.256805,Franke2019}. Therefore, in the presence of impurity chains and lattices the coherent overlapping impurity states can give rise to a long-range hopping models and topological superconductivity \cite{Pientka13, pientka2014, Rontynen15}. The one-dimensional magnetic adatom chains are theoretically predicted to support Majorana zero modes localized at the end of the chain \cite{Choy11, Braunecker2013, klinovaja2013, vazifeh2013, Pientka13, pientka2014,PhysRevLett.128.036801}
and promising signatures  have been experimentally observed \cite{Nad14, ruby2015, Kim18, schneider2021, schneider2021a}. These observations have attracted significant interest in the quantum computing community because the non-Abelian braiding statistics of the Majorana zero modes. In particular, they can be utilized in topological quantum computing and the quantum information stored in the Majorana qubits is protected from local sources of noise \cite{Nay08,Beenakker20}. Although most of the braiding proposals have been developed for Majorana nanowires \cite{Beenakker20},  different methods for manipulating the quantum information in Yu-Shiba-Rusinov chains are also currently being explored \cite{Li2016, Kreisel21, mishra2021yushibarusinov}.

The case of a two-dimensional Yu-Shiba-Rusinov lattice is theoretically even more interesting \cite{nakosai2013, Kimme2015, Rontynen15,  Bernevig16, Kimme16, Teemu16,2022arXiv220211003K,superYSR2020,PhysRevB.96.184425,PhysRevLett.120.017001}. In this case the hybridization of the Yu-Shiba-Rusinov states gives rise a rich phase diagram containing large number of different topological phases \cite{Rontynen15,Bernevig16, Kimme16, Teemu16}. These topological phases are described by an integer-valued Chern number $C$ \cite{TKNN}, which determines the number of chiral Majorana edge modes (see Fig.~\ref{fig:fig0}) and value of the quantized thermal conductance \cite{ReadGreen, SenthilFisher}. So far the quantization of the thermal conductance in superconductors has remained elusive, but recent experiments show promising signatures of the Majorana edge modes in the local density of states (LDOS) data measured via the scanning tunneling spectroscopy techniques \cite{Trif17, Wiesendager19, kezilebieke2020topological, Kezilebieke22}, and unbiased methods for identifying the Chern number from LDOS are currently under development \cite{PhysRevB.97.115453,MLphase2019,PhysRevB.102.054107,PhysRevLett.125.127401,baireuther2021}. Therefore, there are reasons to be optimistic that the topological phase diagram of the Yu-Shiba-Rusinov lattice system can be probed also experimentally in the near future.

The rich topological phase diagram of the Yu-Shiba-Rusinov lattice model \cite{Rontynen15,Bernevig16, Kimme16, Teemu16} calls for a detailed theoretical analysis of the different factors that may play an important role in the selection of the topological phases, which are actually realized in the experiments. So far most of the studies have mainly focused on extrinsic factors and it is known for example that in the presence of strong disorder only the $C=0$ and $C=1$ phases survive \cite{Kim2018}, out of the dozens of topologically distinct phases appearing in a clean system \cite{Rontynen15, Teemu16}. Here, we study an important intrinsic factor which is present also in the highest quality samples: the effect of electron-electron interactions. We may expect  that they lead to competing effects. On one hand, the interactions favor states with large energy gaps in order to lower the free energy in the many-particle systems. On the other hand, the phases with large Chern numbers are most fragile to the effects of the perturbations because they arise from a complex interplay of the superconductivity, magnetism and large spatial extent of the Yu-Shiba-Rusinov states. In particular, the increase of the energy gap leads to a shorter coherence length decreasing the long-range coupling between the Shiba states which is essential for realizing large Chern numbers.  Therefore, the interactions could favor the phases with small Chern numbers, where it is easier to open sizeable energy gaps. But, in order that the interactions could result in a change of the Chern number from a large to a small value the system, in the absence of first order transitions, has to undergo a series of energy gap closings. Thus, the interactions can also lead to lowering of the energy gaps. We show that these competing tendencies in this type of complicated system supporting dozens of topologically distinct phases  lead to rather surprising statistical effects. By sampling a large number of realizations of Yu-Shiba-Rusinov lattice models with different model parameters, we show that statistically the interactions have practically no effect on the realized Chern numbers and typical magnitudes of the topological gaps. 
However, the interactions substantially enhance the likelihood of the rare realizations of large topological gaps in the tails of the energy gap distribution in comparison to the non-interacting case.


\section{Model}

The Hamiltonian of the system takes the form 
\begin{equation}
    \mathcal{H} = \mathcal{H}_{\text{kin}} + \mathcal{H}_{R} + \mathcal{H}_{J} + \mathcal{H}_{\text{SC}} \label{non-interacting-H}
\end{equation}
where $\mathcal{H}_{\text{kin}}$ describes the nearest-neighbor hopping
\begin{equation}
    \mathcal{H}_{\text{kin}} = t \sum_{\langle ij \rangle} c^\dagger_{i,s} c_{j,s},
\end{equation}
$\mathcal{H}_{R}$ is the Rashba spin-orbit coupling
\begin{equation}
    \mathcal{H}_{R} = i \lambda_R \sum_{\langle ij \rangle} \mathbf{d}_{ij} \cdot \boldsymbol{\sigma}_{s,s'} c^\dagger_{i,s} c_{j,s'},
\end{equation}
$\mathcal{H}_{J}$ is the magnetic exchange coupling induced by the impurities 
\begin{equation}
    \mathcal{H}_{J} = J \sum_{i \in {\rm imp}} \sigma^z_{s,s'} c^\dagger_{i,s} c_{i,s'},
\end{equation}
and $\mathcal{H}_{\text{SC}}$ describes the superconducting pairing
\begin{equation}
    \mathcal{H}_{\text{SC}} = \Delta \sum_{i} c^\dagger_{i,\uparrow} c^\dagger_{i,\downarrow} + \text{h.c.} \label{induced-SC}
\end{equation}
Here $c^\dagger_{i,s}$ is the creation operator for an electron in site $i$ and spin $s$, $\mathbf{d}_{ij} = \mathbf{r}_i - \mathbf{r}_j$, and $i \in {\rm imp}$ indicates that the summation is over the impurity sites. We assume that the impurity sites form a square lattice with lattice constant of $a$, as shown in Fig.~\ref{fig:fig0}(a).

\begin{figure*}
    \centering
    \includegraphics[scale=0.19]{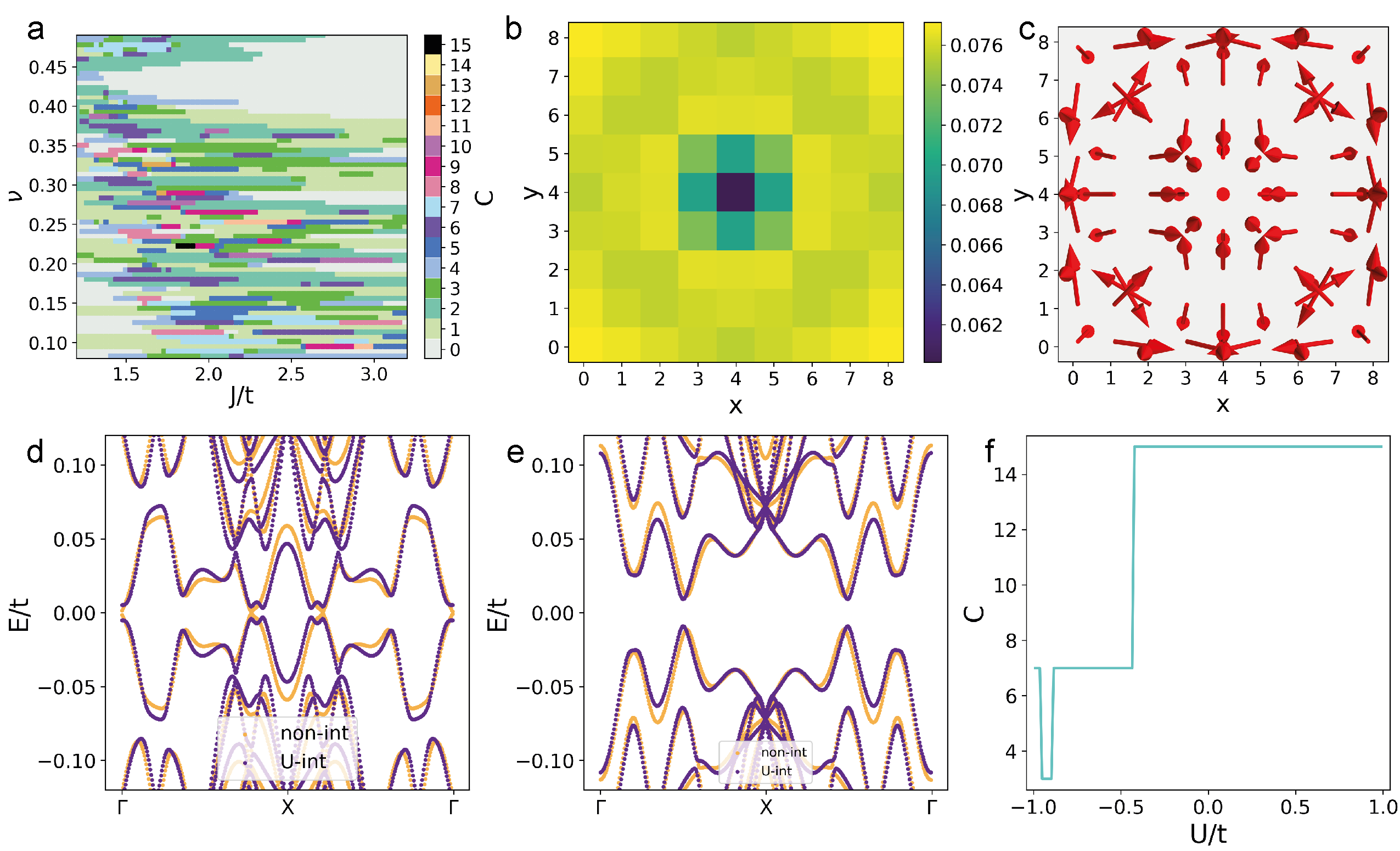}
	\caption{(a) Phase diagram of the non-interacting model as a function of filling $\nu$ and the exchange interaction $J$ for $\alpha=0.1 t$ and $\Delta=0.06 t$.   (b), (c) Illustrations of effects of interactions on the mean field superconductivity and magnetization, respectively. The magnetic impurity is located at the center of the unit cell. The parameters are (b) $\nu=0.101$, $J=2.433 t$, $\alpha=0.112 t$, $\Delta=0.054 t$, $U=-0.702 t$ and (c) $\nu=0.115$, $J=2.172 t$, $\alpha=0.117 t$, $\Delta=0.052 t$, $U=0.644 t$, $V=0.012 t$.  (d), (e) Examples of spectra where the interactions enhance the gap and reduce the gap, respectively. The parameters are  (d) $\nu=0.115$, $J=2.172 t$, $\alpha=0.117 t$, $\Delta=0.052 t$, $U=0.644 t$, $V=0.012 t$  and  (e) $\nu=0.1$, $J=2.272 t$, $\alpha=0.113 t$, $\Delta=0.06 t$, $U=-0.37 t$. (f) Illustration of topological phase transitions as a function of interaction strength $U$ for $\nu=0.1$, $J=2.272t$, $\alpha=0.113t$, $\Delta=0.06$. In all cases we have asssumed a square lattice of impurities with a lattice constant $a=8$.
}
\label{fig:fig1}
\end{figure*}

The many-body interactions are included by means of an local $U$ and nearest-neighbor $V$ density-density interaction [see Fig.~\ref{fig:fig0}(b)] of the form
\begin{equation}
    \mathcal{H}_{\text{int}} = \mathcal{H}_U + \mathcal{H}_V
\end{equation}
\begin{equation}
    \mathcal{H}_U = U \sum_i c^\dagger_{i,\uparrow} c_{i,\uparrow} c^\dagger_{i,\downarrow} c_{i,\downarrow}
\end{equation}
\begin{equation}
    \mathcal{H}_V = V \sum_{\langle i j \rangle} 
    \left ( \sum_s c^\dagger_{i,s} c_{i,s} \right )
    \left ( \sum_s c^\dagger_{j,s} c_{j,s} \right )
\end{equation}
The Hamiltonian is solved at the mean-field level
$\mathcal{H}_{\text{int}} \approx \mathcal{H}_{\text{int}}^{MF}$
including all the bilinear contractions of the mean-field
\begin{equation}
\mathcal{H}_{\text{int}}^{MF} =
    \sum_{i,j,s,s'} \chi^{s,s'}_{i,j} c^\dagger_{i,s}c_{j,s'} +
    \sum_{i,j,s,s'} \xi^{s,s'}_{i,j} c^\dagger_{i,s}c^\dagger_{j,s'} + \text{h.c.}
\end{equation}
giving rise to hopping, pairing, chemical potential and spin-orbit coupling renormalization \cite{pyqula}. By studying separately the cases where only $U$ is present ($U$-interacting case) and both $U$ and $V$ are present ($UV$-interacting case) we can also get some idea how the range of the interactions affects the results. The self-consistent
calculations are performed at fixed filling of the normal state $\nu \in [0,1]$.  Since we assume everywhere that $\Delta \ll t$,  $\nu$ also approximately describes the filling of the bands in the presence of superconductivity. We compute the Chern number numerically using a gauge-invariant description of Chern number associated with the Berry connection defined on a discretized Brillouin zone \cite{Fukui05}. Importantly, because the sign of the Chern number is not important for our analysis, everywhere in the manuscript we denote with $C$ the absolute value of the Chern number. 

For concreteness, we show in Fig.~\ref{fig:fig0}(c) the electronic structure of a specific example of the Yu-Shiba-Rusinov lattice model. The corresponding Berry curvature distribution $\Omega(\mathbf{k})$ is shown in
Fig.~\ref{fig:fig0}(d), and by integrating $\Omega(\mathbf{k})$ over the Brillouin zone we obtain a Chern number $C=7$.
As a consequence of $C=7$, the spectral function in a semi-infinite geometry
features 7 chiral edge modes crossing the energy gap, as shown in Fig.~\ref{fig:fig0}(e).

\section{Effects of interactions on individual realizations}

In the limit when $\Delta \ll t$ and the impurity lattice constant satisfies $a \gg 1$, the non-interacting model of Eqs.~(\ref{non-interacting-H})-(\ref{induced-SC}) supports a rich topological phase diagram as a function of model parameters \cite{Rontynen15, Bernevig16, Teemu16}. In Fig.~\ref{fig:fig1}(a) we show a representative phase diagram for $a=8$ as a function of filling $\nu$ and the exchange interaction $J$ for $\alpha=0.1 t$ and $\Delta=0.06 t$.  It contains topological phases with Chern numbers ranging from $0$ to $15$ arising due to the long-range coupling between  
the Yu-Shiba-Rusinov states \cite{Rontynen15}. 
The high Chern number topological phases originate from rather peculiar situations where the longer range couplings between the Shiba states are larger than the shorter range couplings, and therefore they cover only small areas of the parameter space [see Fig.~\ref{fig:fig1}(a)].  This also means that the phases with large $C$ are quite fragile against the effects of perturbations and for example in the presence of strong disorder only the $C=0$ and $C=1$ phases survive \cite{Kim2018}. 

The electron-electron interactions can influence the topological phases because they can modify the spatial profiles of the mean fields. We find that in the presence of interactions the superconducting mean field is modulated with the period $a$ of the magnetic impurities. A representative example of a single unit cell of the impurity lattice with the magnetic impurity located at the center of it is shown in Fig.~\ref{fig:fig1}(b). In the case of magnetization, the interactions do not only modify the magnitude of the mean field, but also lead to an appearance of magnetization textures where the direction of the magnetization varies within the unit cell [see Fig.~\ref{fig:fig1}(c)]. 
We find that these effects can both enhance and reduce the topological energy gap, and even cause topological phase transitions [see Figs.~\ref{fig:fig1}(d)-(f)]. Therefore, we conclude that the interaction effects can play an important role in the determination of the topological phase and the magnitude of the topological gap  in the case of individual samples.

\section{Statistical effects of interactions}

\begin{figure}
    \centering
    \includegraphics[scale=0.19]{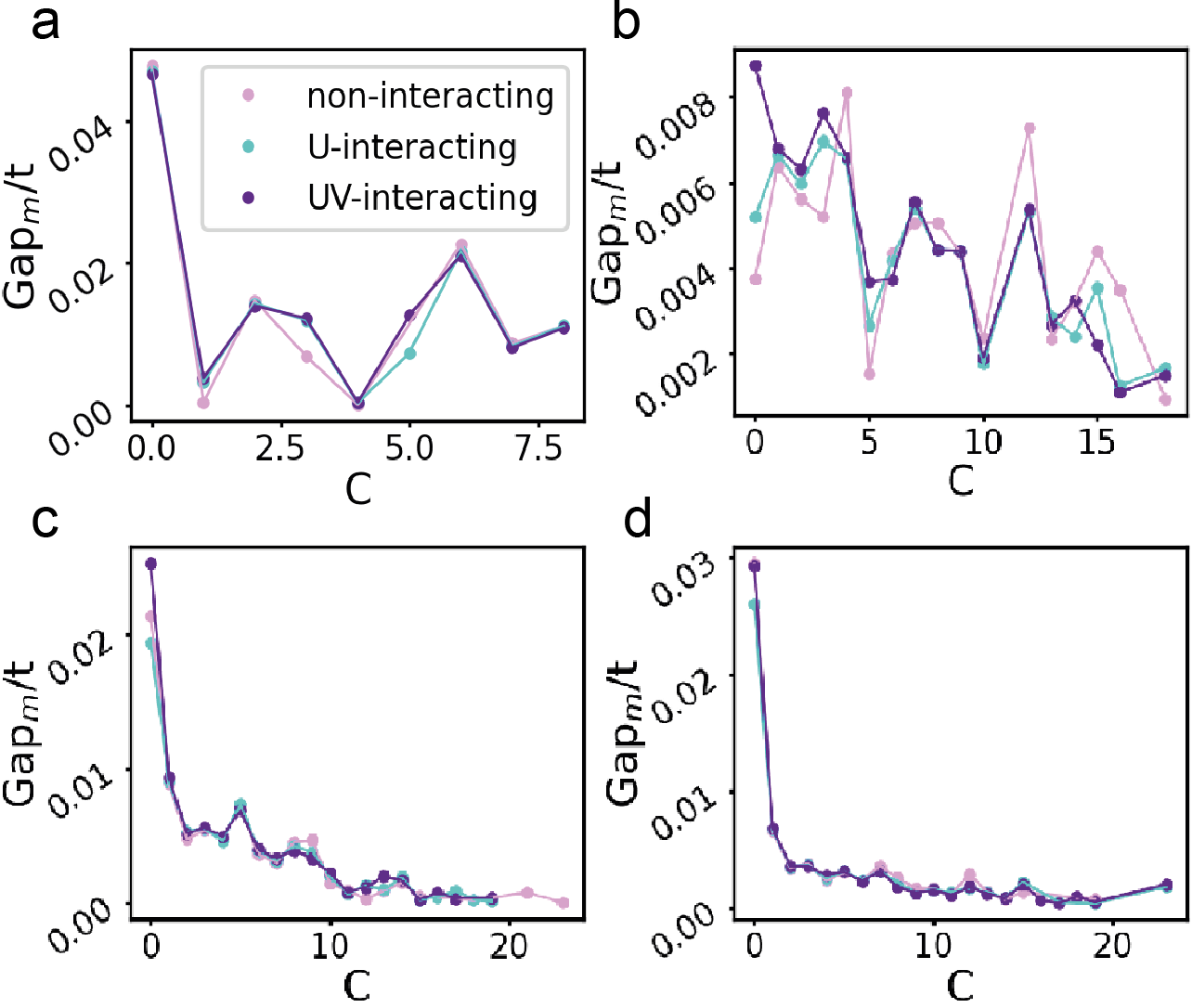}
	\caption{Median energy gap values for different values of $C$ in the non-interacting, $U$-interacting, $UV$-interacting cases for different impurity lattice constants (a) $a=2$, (b) $a=4$, (c) $a=6$ and (d) $a=8$. The sampling of the model parameters is described in the main text. 
}
\label{fig:fig2}
\end{figure}

\begin{figure*}
    \centering
    \includegraphics[scale=0.19]{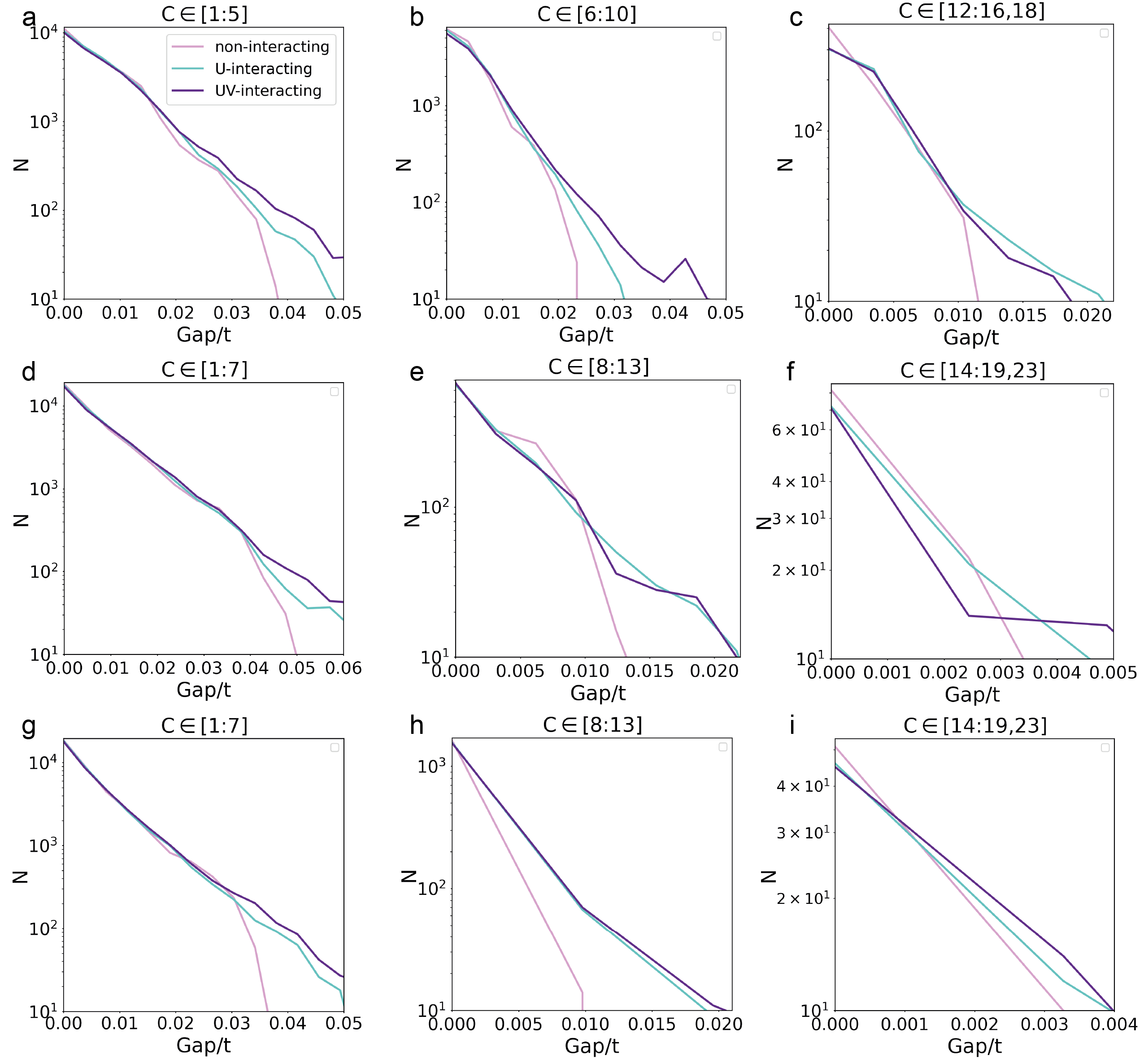}
	\caption{
	Statistical distributions of the energy gap values in the non-interacting, $U$-interacting and $UV$-interacting cases for different lattice constants (a)-(c) $a=4$, (d)-(f) $a=6$ and (g)-(i) $a=8$, respectively. The statistical distributions are nearly identical in all cases in the regime of small energy gaps. In contrast, interactions substantially modify the tails of the distributions increasing the likelihood of large topological gaps.
}
\label{fig:fig3}
\end{figure*}

The model parameters are rarely known accurately in experiments. Therefore, rather than studying the system for specific model parameters, it is important to establish the statistical effects of the interactions on the existence of the topological phases with high $C$ and for the magnitudes of the corresponding energy gaps. 
For this purpose we have collected statistics of the gap sizes and Chern numbers for 50000 sets of model parameters  in impurity lattices with lattice constant $a=:2, 4, 6, 8$. In the sampling we used common ranges   $\nu \in [0.1, 0.45]$, $\alpha \in t[0.08, 0.12]$, $\Delta \in t[0.048, 0.072]$, $U\in t[-1.0, 1.0]$ and $V\in t[-0.5, 0.5]$, whereas the sampling of $J$ for each $a$ was chosen so   that the topological phase diagrams contained a large number of topologically distinct phases. We used $J\in t[0.76, 1.2]$ for $a=2$, $J\in t[1.4,2.1]$ for $a=4$, and $J\in t[1.8,2.7]$ for $a=6$ and $a=8$.

In Fig.~\ref{fig:fig2} we show the median energy gap values for different values of $C$ in the non-interacting, $U$-interacting, $UV$-interacting cases for different impurity lattice constants. The median gap values oscillate as a function of $C$ but there is also a general tendency for the energy gaps to become smaller and smaller with increasing values of $C$. Importantly, we find that the median energy gaps have remarkably similar magnitudes and dependence on $C$ in the non-interacting, $U$-interacting and $UV$-interacting cases (see Fig.~\ref{fig:fig2}). This suggests that the interactions are not statistically relevant in the determination of the topological phases and magnitudes of the topological gaps within the quite large range of interaction strengths considered. We believe that the interval of the   electron-electron interaction strengths is sufficiently large to draw reliable conclusions, because the interactions are screened by the superconductor and therefore larger interaction strengths are unlikely to occur in the realization of the Shiba lattice models.     

The median gap values of course give only partial information about the distribution of the gap values, and it could be possible that the interactions still affect the shape of this distribution. In Fig.~\ref{fig:fig3} we plot the whole statistical distribution of the gap values in the cases of small, intermediate and large Chern numbers. The distributions are practically identical in the non-interacting, $U$-interacting and $UV$-interacting cases in the case of small gap sizes. Since the distribution function decays approximately exponentially with the increasing energy gap values (see Fig.~\ref{fig:fig3}) the small gap values occur much more commonly than the large ones, so that the typical gap sizes are indeed the same in the interacting and non-interacting cases. However, the rare events corresponding to large topological
gaps show a drastically different phenomenology. As shown in Fig.~\ref{fig:fig3}, for all the
Chern numbers interactions substantially increase the probability of the largest topological
gaps. In that regime, the non-interacting gap distribution decreases much faster than exponentially when the energy gap values approach the upper bound of the gap values of the corresponding Chern numbers that can be realized in the non-interacting Yu-Shiba-Rusinov model. In the presence of interactions, the number of samples exhibiting a large topological gap becomes much larger than in the non-interacting case, demonstrating
that the interactions play an important role in the case of these rare events.

\section{Effects beyond mean-field}
So far, our analysis has focused on a purely mean-field description of electronic interactions. Effects beyond mean-field are well known to be important for two-dimensional systems,
and could give rise to several different types of  corrections. 
 First, at the mean-field level, the electronic states have effectively an infinite lifetime. However, 
in  interacting systems the presence of interactions can give rise to
finite lifetimes even for low-lying modes \cite{FL2018}.
Furthermore, quantum fluctuations are not accounted for
in the mean-field methodology, resulting in an overestimation
of the symmetry breaking effects \cite{Hubbard2015}. 
It is also worth noting that
our analysis treats the magnetic impurity as a classical spin, neglecting quantum fluctuation effects.
In reality, such quantum fluctuations can give rise  to a crossover between Kondo and Yu-Shiba-Rusinov physics \cite{PhysRevLett.127.186804}, which by construction are not
included in our model. Finally, the effects of the interactions beyond the mean-field theory in the considered high-Chern number quasi-flat-band system could also lead to qualitatively new physics such as  fractionalization, leading to more exotic non-Abelian phases.   

It is interesting to note
that, while a variety of methods exist for interacting two-dimensional models, the current problem
would provide a great challenge for most of them. Exact diagonalization provides an accurate methodology
to approach small clusters. Nonetheless, due to the large number of orbitals per unit cell,
including large number of unit cells in an exact diagonalization would not be feasible. It is
worth noting that such a limitation also emerges in moir\'e systems, where supercells contains many sites.
In moir\'e systems, however, projection of interactions on a single or few low-energy bands allows
to get a reduced effective model per supercell \cite{PhysRevLett.124.106803,PhysRevResearch.2.023238}. In the Yu-Shiba-Rusinov model considered, the
different bands are often highly entangled, and in such kind of situations isolating a single low energy band is not feasible.
Tensor-network \cite{Orus2014} and neural-network quantum state methodologies \cite{Carleo2017} have become one of the more powerful
unbiased methods to treat two-dimensional systems. However, again due to the large size of the supercell, these calculations would become unfeasible in practice.
An alternative strategy would be to use a dynamical mean-field theory \cite{RevModPhys.68.13}
for the supercell considered,
which would be possible for modest supercells using the current state-of-the art methodology.
The dynamical mean-field theory would allow to obtain
a more accurate picture of the renormalization
of electronic structure due to interactions, and more importantly, to account for additional   
Kondo physics.
Finally, GW methodology of the effective model \cite{RevModPhys.74.601} in the normal state
can yield a more accurate picture of the renormalization effects on the
electronic structure by electronic repulsion. 
While the study of the interaction effects beyond the mean-field level is certainly an interesting direction for future research, the interactions in our calculations do not lead to spontaneous breaking of symmetries. Thus, the mean-field calculations can provide a reliable description of the many-body states also in low-dimensional systems \cite{Stoudenmire}.

\section{Conclusions}

To conclude, we have studied  the effects of electron-electron interactions in the two-dimensional Yu-Shiba-Rusinov lattice model on mean-field level, and our results show that the interactions can enhance or reduce the topological gap as well as cause topological phase transitions. We have shown that statistically the distributions of the Chern numbers and the topological energy gaps in the non-interacting and interacting cases are approximately the same for the most common realizations where the energy gaps are reasonably small. Interestingly, we find that the probability of rare realizations with large energy gaps, located at the 
tails of statistical distributions, are strongly enhanced by interactions.
Such realization are  challenging to find experimentally due to their small likelihood, but they would be the most suitable ones for the applications of topological superconductivity. Our results provide also a starting point for studies of more subtle correlation effects in Shiba lattice models. The effects beyond the mean-field theory are a great challenge for future theoretical research, but this research direction might be worth pursuing because these models may support unprecedented many-particle phenomenology thanks to the interplay of superconductivity, magnetism, high Chern numbers and the large spatial extent of the Yu-Shiba-Rusinov states.

\begin{acknowledgments}
\textit{Acknowledgments:}
We acknowledge
the computational resources provided by
the Aalto Science-IT project.
We acknowledge
financial support from the
Academy of Finland Projects No.
331094, No.
331342 and No. 336243,
and the Jane and Aatos Erkko Foundation.
\end{acknowledgments}

\bibliography{biblio}

\begin{thebibliography}{61}%
\makeatletter
\providecommand \@ifxundefined [1]{%
 \@ifx{#1\undefined}
}%
\providecommand \@ifnum [1]{%
 \ifnum #1\expandafter \@firstoftwo
 \else \expandafter \@secondoftwo
 \fi
}%
\providecommand \@ifx [1]{%
 \ifx #1\expandafter \@firstoftwo
 \else \expandafter \@secondoftwo
 \fi
}%
\providecommand \natexlab [1]{#1}%
\providecommand \enquote  [1]{``#1''}%
\providecommand \bibnamefont  [1]{#1}%
\providecommand \bibfnamefont [1]{#1}%
\providecommand \citenamefont [1]{#1}%
\providecommand \href@noop [0]{\@secondoftwo}%
\providecommand \href [0]{\begingroup \@sanitize@url \@href}%
\providecommand \@href[1]{\@@startlink{#1}\@@href}%
\providecommand \@@href[1]{\endgroup#1\@@endlink}%
\providecommand \@sanitize@url [0]{\catcode `\\12\catcode `\$12\catcode
  `\&12\catcode `\#12\catcode `\^12\catcode `\_12\catcode `\%12\relax}%
\providecommand \@@startlink[1]{}%
\providecommand \@@endlink[0]{}%
\providecommand \url  [0]{\begingroup\@sanitize@url \@url }%
\providecommand \@url [1]{\endgroup\@href {#1}{\urlprefix }}%
\providecommand \urlprefix  [0]{URL }%
\providecommand \Eprint [0]{\href }%
\providecommand \doibase [0]{https://doi.org/}%
\providecommand \selectlanguage [0]{\@gobble}%
\providecommand \bibinfo  [0]{\@secondoftwo}%
\providecommand \bibfield  [0]{\@secondoftwo}%
\providecommand \translation [1]{[#1]}%
\providecommand \BibitemOpen [0]{}%
\providecommand \bibitemStop [0]{}%
\providecommand \bibitemNoStop [0]{.\EOS\space}%
\providecommand \EOS [0]{\spacefactor3000\relax}%
\providecommand \BibitemShut  [1]{\csname bibitem#1\endcsname}%
\let\auto@bib@innerbib\@empty
\bibitem [{\citenamefont {Yu}(1965)}]{yu1965}%
  \BibitemOpen
  \bibfield  {author} {\bibinfo {author} {\bibfnamefont {L.}~\bibnamefont
  {Yu}},\ }\bibfield  {title} {\bibinfo {title} {{Bound state in
  superconductors with paramagnetic impurities}},\ }\href@noop {} {\bibfield
  {journal} {\bibinfo  {journal} {Acta Phys. Sin}\ }\textbf {\bibinfo {volume}
  {21}},\ \bibinfo {pages} {75} (\bibinfo {year} {1965})}\BibitemShut {NoStop}%
\bibitem [{\citenamefont {Shiba}(1968)}]{shiba1968}%
  \BibitemOpen
  \bibfield  {author} {\bibinfo {author} {\bibfnamefont {H.}~\bibnamefont
  {Shiba}},\ }\bibfield  {title} {\bibinfo {title} {{Classical spins in
  superconductors}},\ }\href@noop {} {\bibfield  {journal} {\bibinfo  {journal}
  {Progress of Theoretical Physics}\ }\textbf {\bibinfo {volume} {40}},\
  \bibinfo {pages} {435} (\bibinfo {year} {1968})}\BibitemShut {NoStop}%
\bibitem [{\citenamefont {Rusinov}(1969)}]{rusinov1969}%
  \BibitemOpen
  \bibfield  {author} {\bibinfo {author} {\bibfnamefont {A.}~\bibnamefont
  {Rusinov}},\ }\bibfield  {title} {\bibinfo {title} {{Superconductivity near a
  paramagnetic impurity}},\ }\href@noop {} {\bibfield  {journal} {\bibinfo
  {journal} {Soviet Journal of Experimental and Theoretical Physics Letters}\
  }\textbf {\bibinfo {volume} {9}},\ \bibinfo {pages} {85} (\bibinfo {year}
  {1969})}\BibitemShut {NoStop}%
\bibitem [{\citenamefont {Balatsky}\ \emph {et~al.}(2006)\citenamefont
  {Balatsky}, \citenamefont {Vekhter},\ and\ \citenamefont
  {Zhu}}]{BalatskyRMP06}%
  \BibitemOpen
  \bibfield  {author} {\bibinfo {author} {\bibfnamefont {A.~V.}\ \bibnamefont
  {Balatsky}}, \bibinfo {author} {\bibfnamefont {I.}~\bibnamefont {Vekhter}},\
  and\ \bibinfo {author} {\bibfnamefont {J.-X.}\ \bibnamefont {Zhu}},\
  }\bibfield  {title} {\bibinfo {title} {{Impurity-induced states in
  conventional and unconventional superconductors}},\ }\href
  {https://doi.org/10.1103/RevModPhys.78.373} {\bibfield  {journal} {\bibinfo
  {journal} {Rev. Mod. Phys.}\ }\textbf {\bibinfo {volume} {78}},\ \bibinfo
  {pages} {373} (\bibinfo {year} {2006})}\BibitemShut {NoStop}%
\bibitem [{\citenamefont {M{\'e}nard}\ \emph {et~al.}(2015)\citenamefont
  {M{\'e}nard}, \citenamefont {Guissart}, \citenamefont {Brun}, \citenamefont
  {Pons}, \citenamefont {Stolyarov}, \citenamefont {Debontridder},
  \citenamefont {Leclerc}, \citenamefont {Janod}, \citenamefont {Cario},
  \citenamefont {Roditchev} \emph {et~al.}}]{menard2015}%
  \BibitemOpen
  \bibfield  {author} {\bibinfo {author} {\bibfnamefont {G.~C.}\ \bibnamefont
  {M{\'e}nard}}, \bibinfo {author} {\bibfnamefont {S.}~\bibnamefont
  {Guissart}}, \bibinfo {author} {\bibfnamefont {C.}~\bibnamefont {Brun}},
  \bibinfo {author} {\bibfnamefont {S.}~\bibnamefont {Pons}}, \bibinfo {author}
  {\bibfnamefont {V.~S.}\ \bibnamefont {Stolyarov}}, \bibinfo {author}
  {\bibfnamefont {F.}~\bibnamefont {Debontridder}}, \bibinfo {author}
  {\bibfnamefont {M.~V.}\ \bibnamefont {Leclerc}}, \bibinfo {author}
  {\bibfnamefont {E.}~\bibnamefont {Janod}}, \bibinfo {author} {\bibfnamefont
  {L.}~\bibnamefont {Cario}}, \bibinfo {author} {\bibfnamefont
  {D.}~\bibnamefont {Roditchev}}, \emph {et~al.},\ }\bibfield  {title}
  {\bibinfo {title} {{Coherent long-range magnetic bound states in a
  superconductor}},\ }\href {https://doi.org/10.1038/nphys3508} {\bibfield
  {journal} {\bibinfo  {journal} {Nature Physics}\ }\textbf {\bibinfo {volume}
  {11}},\ \bibinfo {pages} {1013} (\bibinfo {year} {2015})}\BibitemShut
  {NoStop}%
\bibitem [{\citenamefont {Cortés‐del~Río}\ \emph
  {et~al.}(2021)\citenamefont {Cortés‐del~Río}, \citenamefont {Lado},
  \citenamefont {Cherkez}, \citenamefont {Mallet}, \citenamefont {Veuillen},
  \citenamefont {Cuevas}, \citenamefont {Gómez‐Rodríguez}, \citenamefont
  {Fernández‐Rossier},\ and\ \citenamefont {Brihuega}}]{grapheneYSR2021}%
  \BibitemOpen
  \bibfield  {author} {\bibinfo {author} {\bibfnamefont {E.}~\bibnamefont
  {Cortés‐del~Río}}, \bibinfo {author} {\bibfnamefont {J.~L.}\ \bibnamefont
  {Lado}}, \bibinfo {author} {\bibfnamefont {V.}~\bibnamefont {Cherkez}},
  \bibinfo {author} {\bibfnamefont {P.}~\bibnamefont {Mallet}}, \bibinfo
  {author} {\bibfnamefont {J.}~\bibnamefont {Veuillen}}, \bibinfo {author}
  {\bibfnamefont {J.~C.}\ \bibnamefont {Cuevas}}, \bibinfo {author}
  {\bibfnamefont {J.~M.}\ \bibnamefont {Gómez‐Rodríguez}}, \bibinfo
  {author} {\bibfnamefont {J.}~\bibnamefont {Fernández‐Rossier}},\ and\
  \bibinfo {author} {\bibfnamefont {I.}~\bibnamefont {Brihuega}},\ }\bibfield
  {title} {\bibinfo {title} {{Observation of Yu–Shiba–Rusinov States in
  Superconducting Graphene}},\ }\href {https://doi.org/10.1002/adma.202008113}
  {\bibfield  {journal} {\bibinfo  {journal} {Advanced Materials}\ }\textbf
  {\bibinfo {volume} {33}},\ \bibinfo {pages} {2008113} (\bibinfo {year}
  {2021})}\BibitemShut {NoStop}%
\bibitem [{\citenamefont {Kezilebieke}\ \emph {et~al.}(2018)\citenamefont
  {Kezilebieke}, \citenamefont {Dvorak}, \citenamefont {Ojanen},\ and\
  \citenamefont {Liljeroth}}]{KeziYSR2018}%
  \BibitemOpen
  \bibfield  {author} {\bibinfo {author} {\bibfnamefont {S.}~\bibnamefont
  {Kezilebieke}}, \bibinfo {author} {\bibfnamefont {M.}~\bibnamefont {Dvorak}},
  \bibinfo {author} {\bibfnamefont {T.}~\bibnamefont {Ojanen}},\ and\ \bibinfo
  {author} {\bibfnamefont {P.}~\bibnamefont {Liljeroth}},\ }\bibfield  {title}
  {\bibinfo {title} {{Coupled Yu–Shiba–Rusinov States in Molecular Dimers
  on NbSe$_2$}},\ }\href {https://doi.org/10.1021/acs.nanolett.7b05050}
  {\bibfield  {journal} {\bibinfo  {journal} {Nano Letters}\ }\textbf {\bibinfo
  {volume} {18}},\ \bibinfo {pages} {2311–2315} (\bibinfo {year}
  {2018})}\BibitemShut {NoStop}%
\bibitem [{\citenamefont {Farinacci}\ \emph {et~al.}(2020)\citenamefont
  {Farinacci}, \citenamefont {Ahmadi}, \citenamefont {Ruby}, \citenamefont
  {Reecht}, \citenamefont {Heinrich}, \citenamefont {Czekelius}, \citenamefont
  {von Oppen},\ and\ \citenamefont {Franke}}]{PhysRevLett.125.256805}%
  \BibitemOpen
  \bibfield  {author} {\bibinfo {author} {\bibfnamefont {L.}~\bibnamefont
  {Farinacci}}, \bibinfo {author} {\bibfnamefont {G.}~\bibnamefont {Ahmadi}},
  \bibinfo {author} {\bibfnamefont {M.}~\bibnamefont {Ruby}}, \bibinfo {author}
  {\bibfnamefont {G.}~\bibnamefont {Reecht}}, \bibinfo {author} {\bibfnamefont
  {B.~W.}\ \bibnamefont {Heinrich}}, \bibinfo {author} {\bibfnamefont
  {C.}~\bibnamefont {Czekelius}}, \bibinfo {author} {\bibfnamefont
  {F.}~\bibnamefont {von Oppen}},\ and\ \bibinfo {author} {\bibfnamefont
  {K.~J.}\ \bibnamefont {Franke}},\ }\bibfield  {title} {\bibinfo {title}
  {{Interfering Tunneling Paths through Magnetic Molecules on Superconductors:
  Asymmetries of Kondo and Yu-Shiba-Rusinov Resonances}},\ }\href
  {https://doi.org/10.1103/PhysRevLett.125.256805} {\bibfield  {journal}
  {\bibinfo  {journal} {Phys. Rev. Lett.}\ }\textbf {\bibinfo {volume} {125}},\
  \bibinfo {pages} {256805} (\bibinfo {year} {2020})}\BibitemShut {NoStop}%
\bibitem [{\citenamefont {Liebhaber}\ \emph {et~al.}(2019)\citenamefont
  {Liebhaber}, \citenamefont {Acero~González}, \citenamefont {Baba},
  \citenamefont {Reecht}, \citenamefont {Heinrich}, \citenamefont {Rohlf},
  \citenamefont {Rossnagel}, \citenamefont {von Oppen},\ and\ \citenamefont
  {Franke}}]{Franke2019}%
  \BibitemOpen
  \bibfield  {author} {\bibinfo {author} {\bibfnamefont {E.}~\bibnamefont
  {Liebhaber}}, \bibinfo {author} {\bibfnamefont {S.}~\bibnamefont
  {Acero~González}}, \bibinfo {author} {\bibfnamefont {R.}~\bibnamefont
  {Baba}}, \bibinfo {author} {\bibfnamefont {G.}~\bibnamefont {Reecht}},
  \bibinfo {author} {\bibfnamefont {B.~W.}\ \bibnamefont {Heinrich}}, \bibinfo
  {author} {\bibfnamefont {S.}~\bibnamefont {Rohlf}}, \bibinfo {author}
  {\bibfnamefont {K.}~\bibnamefont {Rossnagel}}, \bibinfo {author}
  {\bibfnamefont {F.}~\bibnamefont {von Oppen}},\ and\ \bibinfo {author}
  {\bibfnamefont {K.~J.}\ \bibnamefont {Franke}},\ }\bibfield  {title}
  {\bibinfo {title} {{Yu–Shiba–Rusinov States in the Charge-Density
  Modulated Superconductor NbSe$_2$}},\ }\href
  {https://doi.org/10.1021/acs.nanolett.9b03988} {\bibfield  {journal}
  {\bibinfo  {journal} {Nano Letters}\ }\textbf {\bibinfo {volume} {20}},\
  \bibinfo {pages} {339–344} (\bibinfo {year} {2019})}\BibitemShut {NoStop}%
\bibitem [{\citenamefont {Pientka}\ \emph {et~al.}(2013)\citenamefont
  {Pientka}, \citenamefont {Glazman},\ and\ \citenamefont {von
  Oppen}}]{Pientka13}%
  \BibitemOpen
  \bibfield  {author} {\bibinfo {author} {\bibfnamefont {F.}~\bibnamefont
  {Pientka}}, \bibinfo {author} {\bibfnamefont {L.~I.}\ \bibnamefont
  {Glazman}},\ and\ \bibinfo {author} {\bibfnamefont {F.}~\bibnamefont {von
  Oppen}},\ }\bibfield  {title} {\bibinfo {title} {Topological superconducting
  phase in helical {S}hiba chains},\ }\href
  {https://doi.org/10.1103/PhysRevB.88.155420} {\bibfield  {journal} {\bibinfo
  {journal} {Phys. Rev. B}\ }\textbf {\bibinfo {volume} {88}},\ \bibinfo
  {pages} {155420} (\bibinfo {year} {2013})}\BibitemShut {NoStop}%
\bibitem [{\citenamefont {Pientka}\ \emph {et~al.}(2014)\citenamefont
  {Pientka}, \citenamefont {Glazman},\ and\ \citenamefont {von
  Oppen}}]{pientka2014}%
  \BibitemOpen
  \bibfield  {author} {\bibinfo {author} {\bibfnamefont {F.}~\bibnamefont
  {Pientka}}, \bibinfo {author} {\bibfnamefont {L.~I.}\ \bibnamefont
  {Glazman}},\ and\ \bibinfo {author} {\bibfnamefont {F.}~\bibnamefont {von
  Oppen}},\ }\bibfield  {title} {\bibinfo {title} {{Unconventional topological
  phase transitions in helical Shiba chains}},\ }\href
  {https://journals.aps.org/prb/abstract/10.1103/PhysRevB.89.180505} {\bibfield
   {journal} {\bibinfo  {journal} {Physical Review B}\ }\textbf {\bibinfo
  {volume} {89}},\ \bibinfo {pages} {180505} (\bibinfo {year}
  {2014})}\BibitemShut {NoStop}%
\bibitem [{\citenamefont {R\"ontynen}\ and\ \citenamefont
  {Ojanen}(2015)}]{Rontynen15}%
  \BibitemOpen
  \bibfield  {author} {\bibinfo {author} {\bibfnamefont {J.}~\bibnamefont
  {R\"ontynen}}\ and\ \bibinfo {author} {\bibfnamefont {T.}~\bibnamefont
  {Ojanen}},\ }\bibfield  {title} {\bibinfo {title} {{Topological
  Superconductivity and High Chern Numbers in 2D Ferromagnetic Shiba
  Lattices}},\ }\href {https://doi.org/10.1103/PhysRevLett.114.236803}
  {\bibfield  {journal} {\bibinfo  {journal} {Phys. Rev. Lett.}\ }\textbf
  {\bibinfo {volume} {114}},\ \bibinfo {pages} {236803} (\bibinfo {year}
  {2015})}\BibitemShut {NoStop}%
\bibitem [{\citenamefont {Choy}\ \emph {et~al.}(2011)\citenamefont {Choy},
  \citenamefont {Edge}, \citenamefont {Akhmerov},\ and\ \citenamefont
  {Beenakker}}]{Choy11}%
  \BibitemOpen
  \bibfield  {author} {\bibinfo {author} {\bibfnamefont {T.-P.}\ \bibnamefont
  {Choy}}, \bibinfo {author} {\bibfnamefont {J.~M.}\ \bibnamefont {Edge}},
  \bibinfo {author} {\bibfnamefont {A.~R.}\ \bibnamefont {Akhmerov}},\ and\
  \bibinfo {author} {\bibfnamefont {C.~W.~J.}\ \bibnamefont {Beenakker}},\
  }\bibfield  {title} {\bibinfo {title} {Majorana fermions emerging from
  magnetic nanoparticles on a superconductor without spin-orbit coupling},\
  }\href {https://doi.org/10.1103/PhysRevB.84.195442} {\bibfield  {journal}
  {\bibinfo  {journal} {Phys. Rev. B}\ }\textbf {\bibinfo {volume} {84}},\
  \bibinfo {pages} {195442} (\bibinfo {year} {2011})}\BibitemShut {NoStop}%
\bibitem [{\citenamefont {Braunecker}\ and\ \citenamefont
  {Simon}(2013)}]{Braunecker2013}%
  \BibitemOpen
  \bibfield  {author} {\bibinfo {author} {\bibfnamefont {B.}~\bibnamefont
  {Braunecker}}\ and\ \bibinfo {author} {\bibfnamefont {P.}~\bibnamefont
  {Simon}},\ }\bibfield  {title} {\bibinfo {title} {{Interplay between
  Classical Magnetic Moments and Superconductivity in Quantum One-Dimensional
  Conductors: Toward a Self-Sustained Topological {M}ajorana Phase}},\ }\href
  {https://doi.org/10.1103/PhysRevLett.111.147202} {\bibfield  {journal}
  {\bibinfo  {journal} {Phys. Rev. Lett.}\ }\textbf {\bibinfo {volume} {111}},\
  \bibinfo {pages} {147202} (\bibinfo {year} {2013})}\BibitemShut {NoStop}%
\bibitem [{\citenamefont {Klinovaja}\ \emph {et~al.}(2013)\citenamefont
  {Klinovaja}, \citenamefont {Stano}, \citenamefont {Yazdani},\ and\
  \citenamefont {Loss}}]{klinovaja2013}%
  \BibitemOpen
  \bibfield  {author} {\bibinfo {author} {\bibfnamefont {J.}~\bibnamefont
  {Klinovaja}}, \bibinfo {author} {\bibfnamefont {P.}~\bibnamefont {Stano}},
  \bibinfo {author} {\bibfnamefont {A.}~\bibnamefont {Yazdani}},\ and\ \bibinfo
  {author} {\bibfnamefont {D.}~\bibnamefont {Loss}},\ }\bibfield  {title}
  {\bibinfo {title} {{Topological superconductivity and Majorana fermions in
  RKKY systems}},\ }\href
  {https://journals.aps.org/prl/abstract/10.1103/PhysRevLett.111.186805}
  {\bibfield  {journal} {\bibinfo  {journal} {Physical Review Letters}\
  }\textbf {\bibinfo {volume} {111}},\ \bibinfo {pages} {186805} (\bibinfo
  {year} {2013})}\BibitemShut {NoStop}%
\bibitem [{\citenamefont {Vazifeh}\ and\ \citenamefont
  {Franz}(2013)}]{vazifeh2013}%
  \BibitemOpen
  \bibfield  {author} {\bibinfo {author} {\bibfnamefont {M.}~\bibnamefont
  {Vazifeh}}\ and\ \bibinfo {author} {\bibfnamefont {M.}~\bibnamefont
  {Franz}},\ }\bibfield  {title} {\bibinfo {title} {{Self-organized topological
  state with Majorana fermions}},\ }\href
  {https://journals.aps.org/prl/abstract/10.1103/PhysRevLett.111.206802}
  {\bibfield  {journal} {\bibinfo  {journal} {Physical Review Letters}\
  }\textbf {\bibinfo {volume} {111}},\ \bibinfo {pages} {206802} (\bibinfo
  {year} {2013})}\BibitemShut {NoStop}%
\bibitem [{\citenamefont {Steiner}\ \emph {et~al.}(2022)\citenamefont
  {Steiner}, \citenamefont {Mora}, \citenamefont {Franke},\ and\ \citenamefont
  {von Oppen}}]{PhysRevLett.128.036801}%
  \BibitemOpen
  \bibfield  {author} {\bibinfo {author} {\bibfnamefont {J.~F.}\ \bibnamefont
  {Steiner}}, \bibinfo {author} {\bibfnamefont {C.}~\bibnamefont {Mora}},
  \bibinfo {author} {\bibfnamefont {K.~J.}\ \bibnamefont {Franke}},\ and\
  \bibinfo {author} {\bibfnamefont {F.}~\bibnamefont {von Oppen}},\ }\bibfield
  {title} {\bibinfo {title} {{Quantum Magnetism and Topological
  Superconductivity in Yu-Shiba-Rusinov Chains}},\ }\href
  {https://doi.org/10.1103/PhysRevLett.128.036801} {\bibfield  {journal}
  {\bibinfo  {journal} {Phys. Rev. Lett.}\ }\textbf {\bibinfo {volume} {128}},\
  \bibinfo {pages} {036801} (\bibinfo {year} {2022})}\BibitemShut {NoStop}%
\bibitem [{\citenamefont {Nadj-Perge}\ \emph {et~al.}(2014)\citenamefont
  {Nadj-Perge}, \citenamefont {Drozdov}, \citenamefont {Li}, \citenamefont
  {Chen}, \citenamefont {Jeon}, \citenamefont {Seo}, \citenamefont {MacDonald},
  \citenamefont {Bernevig},\ and\ \citenamefont {Yazdani}}]{Nad14}%
  \BibitemOpen
  \bibfield  {author} {\bibinfo {author} {\bibfnamefont {S.}~\bibnamefont
  {Nadj-Perge}}, \bibinfo {author} {\bibfnamefont {I.~K.}\ \bibnamefont
  {Drozdov}}, \bibinfo {author} {\bibfnamefont {J.}~\bibnamefont {Li}},
  \bibinfo {author} {\bibfnamefont {H.}~\bibnamefont {Chen}}, \bibinfo {author}
  {\bibfnamefont {S.}~\bibnamefont {Jeon}}, \bibinfo {author} {\bibfnamefont
  {J.}~\bibnamefont {Seo}}, \bibinfo {author} {\bibfnamefont {A.~H.}\
  \bibnamefont {MacDonald}}, \bibinfo {author} {\bibfnamefont {B.~A.}\
  \bibnamefont {Bernevig}},\ and\ \bibinfo {author} {\bibfnamefont
  {A.}~\bibnamefont {Yazdani}},\ }\bibfield  {title} {\bibinfo {title}
  {Observation of {M}ajorana fermions in ferromagnetic atomic chains on a
  superconductor},\ }\href {https://doi.org/10.1126/science.1259327} {\bibfield
   {journal} {\bibinfo  {journal} {Science}\ }\textbf {\bibinfo {volume}
  {346}},\ \bibinfo {pages} {602} (\bibinfo {year} {2014})}\BibitemShut
  {NoStop}%
\bibitem [{\citenamefont {Ruby}\ \emph {et~al.}(2015)\citenamefont {Ruby},
  \citenamefont {Pientka}, \citenamefont {Peng}, \citenamefont {von Oppen},
  \citenamefont {Heinrich},\ and\ \citenamefont {Franke}}]{ruby2015}%
  \BibitemOpen
  \bibfield  {author} {\bibinfo {author} {\bibfnamefont {M.}~\bibnamefont
  {Ruby}}, \bibinfo {author} {\bibfnamefont {F.}~\bibnamefont {Pientka}},
  \bibinfo {author} {\bibfnamefont {Y.}~\bibnamefont {Peng}}, \bibinfo {author}
  {\bibfnamefont {F.}~\bibnamefont {von Oppen}}, \bibinfo {author}
  {\bibfnamefont {B.~W.}\ \bibnamefont {Heinrich}},\ and\ \bibinfo {author}
  {\bibfnamefont {K.~J.}\ \bibnamefont {Franke}},\ }\bibfield  {title}
  {\bibinfo {title} {{End states and subgap structure in proximity-coupled
  chains of magnetic adatoms}},\ }\href
  {https://link.aps.org/doi/10.1103/PhysRevLett.115.197204} {\bibfield
  {journal} {\bibinfo  {journal} {Physical Review Letters}\ }\textbf {\bibinfo
  {volume} {115}},\ \bibinfo {pages} {197204} (\bibinfo {year}
  {2015})}\BibitemShut {NoStop}%
\bibitem [{\citenamefont {Kim}\ \emph {et~al.}(2018)\citenamefont {Kim},
  \citenamefont {Palacio-Morales}, \citenamefont {Posske}, \citenamefont
  {R{\'o}zsa}, \citenamefont {Palot{\'a}s}, \citenamefont {Szunyogh},
  \citenamefont {Thorwart},\ and\ \citenamefont {Wiesendanger}}]{Kim18}%
  \BibitemOpen
  \bibfield  {author} {\bibinfo {author} {\bibfnamefont {H.}~\bibnamefont
  {Kim}}, \bibinfo {author} {\bibfnamefont {A.}~\bibnamefont
  {Palacio-Morales}}, \bibinfo {author} {\bibfnamefont {T.}~\bibnamefont
  {Posske}}, \bibinfo {author} {\bibfnamefont {L.}~\bibnamefont {R{\'o}zsa}},
  \bibinfo {author} {\bibfnamefont {K.}~\bibnamefont {Palot{\'a}s}}, \bibinfo
  {author} {\bibfnamefont {L.}~\bibnamefont {Szunyogh}}, \bibinfo {author}
  {\bibfnamefont {M.}~\bibnamefont {Thorwart}},\ and\ \bibinfo {author}
  {\bibfnamefont {R.}~\bibnamefont {Wiesendanger}},\ }\bibfield  {title}
  {\bibinfo {title} {Toward tailoring {M}ajorana bound states in artificially
  constructed magnetic atom chains on elemental superconductors},\ }\bibfield
  {journal} {\bibinfo  {journal} {Science Advances}\ }\textbf {\bibinfo
  {volume} {4}},\ \href {https://doi.org/10.1126/sciadv.aar5251}
  {10.1126/sciadv.aar5251} (\bibinfo {year} {2018})\BibitemShut {NoStop}%
\bibitem [{\citenamefont {Schneider}\ \emph {et~al.}(2021)\citenamefont
  {Schneider}, \citenamefont {Beck}, \citenamefont {Posske}, \citenamefont
  {Crawford}, \citenamefont {Mascot}, \citenamefont {Rachel}, \citenamefont
  {Wiesendanger},\ and\ \citenamefont {Wiebe}}]{schneider2021}%
  \BibitemOpen
  \bibfield  {author} {\bibinfo {author} {\bibfnamefont {L.}~\bibnamefont
  {Schneider}}, \bibinfo {author} {\bibfnamefont {P.}~\bibnamefont {Beck}},
  \bibinfo {author} {\bibfnamefont {T.}~\bibnamefont {Posske}}, \bibinfo
  {author} {\bibfnamefont {D.}~\bibnamefont {Crawford}}, \bibinfo {author}
  {\bibfnamefont {E.}~\bibnamefont {Mascot}}, \bibinfo {author} {\bibfnamefont
  {S.}~\bibnamefont {Rachel}}, \bibinfo {author} {\bibfnamefont
  {R.}~\bibnamefont {Wiesendanger}},\ and\ \bibinfo {author} {\bibfnamefont
  {J.}~\bibnamefont {Wiebe}},\ }\bibfield  {title} {\bibinfo {title}
  {{Topological Shiba bands in artificial spin chains on superconductors}},\
  }\href {https://doi.org/10.1038/s41567-021-01234-y} {\bibfield  {journal}
  {\bibinfo  {journal} {Nature Physics}\ }\textbf {\bibinfo {volume} {17}},\
  \bibinfo {pages} {943} (\bibinfo {year} {2021})}\BibitemShut {NoStop}%
\bibitem [{\citenamefont {Schneider}\ \emph {et~al.}(2022)\citenamefont
  {Schneider}, \citenamefont {Beck}, \citenamefont {Neuhaus-Steinmetz},
  \citenamefont {R{\'o}zsa}, \citenamefont {Posske}, \citenamefont {Wiebe},\
  and\ \citenamefont {Wiesendanger}}]{schneider2021a}%
  \BibitemOpen
  \bibfield  {author} {\bibinfo {author} {\bibfnamefont {L.}~\bibnamefont
  {Schneider}}, \bibinfo {author} {\bibfnamefont {P.}~\bibnamefont {Beck}},
  \bibinfo {author} {\bibfnamefont {J.}~\bibnamefont {Neuhaus-Steinmetz}},
  \bibinfo {author} {\bibfnamefont {L.}~\bibnamefont {R{\'o}zsa}}, \bibinfo
  {author} {\bibfnamefont {T.}~\bibnamefont {Posske}}, \bibinfo {author}
  {\bibfnamefont {J.}~\bibnamefont {Wiebe}},\ and\ \bibinfo {author}
  {\bibfnamefont {R.}~\bibnamefont {Wiesendanger}},\ }\bibfield  {title}
  {\bibinfo {title} {{Precursors of Majorana modes and their length-dependent
  energy oscillations probed at both ends of atomic Shiba chains}},\ }\href
  {https://doi.org/10.1038/s41565-022-01078-4} {\bibfield  {journal} {\bibinfo
  {journal} {Nature Nanotechnology}\ }\textbf {\bibinfo {volume} {17}},\
  \bibinfo {pages} {384} (\bibinfo {year} {2022})}\BibitemShut {NoStop}%
\bibitem [{\citenamefont {Nayak}\ \emph {et~al.}(2008)\citenamefont {Nayak},
  \citenamefont {Simon}, \citenamefont {Stern}, \citenamefont {Freedman},\ and\
  \citenamefont {Das~Sarma}}]{Nay08}%
  \BibitemOpen
  \bibfield  {author} {\bibinfo {author} {\bibfnamefont {C.}~\bibnamefont
  {Nayak}}, \bibinfo {author} {\bibfnamefont {S.~H.}\ \bibnamefont {Simon}},
  \bibinfo {author} {\bibfnamefont {A.}~\bibnamefont {Stern}}, \bibinfo
  {author} {\bibfnamefont {M.}~\bibnamefont {Freedman}},\ and\ \bibinfo
  {author} {\bibfnamefont {S.}~\bibnamefont {Das~Sarma}},\ }\bibfield  {title}
  {\bibinfo {title} {Non-{A}belian anyons and topological quantum
  computation},\ }\href {https://doi.org/10.1103/RevModPhys.80.1083} {\bibfield
   {journal} {\bibinfo  {journal} {Rev. Mod. Phys.}\ }\textbf {\bibinfo
  {volume} {80}},\ \bibinfo {pages} {1083} (\bibinfo {year}
  {2008})}\BibitemShut {NoStop}%
\bibitem [{\citenamefont {Beenakker}(2020)}]{Beenakker20}%
  \BibitemOpen
  \bibfield  {author} {\bibinfo {author} {\bibfnamefont {C.~W.~J.}\
  \bibnamefont {Beenakker}},\ }\bibfield  {title} {\bibinfo {title} {{Search
  for non-Abelian Majorana braiding statistics in superconductors}},\ }\href
  {https://doi.org/10.21468/SciPostPhysLectNotes.15} {\bibfield  {journal}
  {\bibinfo  {journal} {SciPost Phys. Lect. Notes}\ ,\ \bibinfo {pages} {15}}
  (\bibinfo {year} {2020})}\BibitemShut {NoStop}%
\bibitem [{\citenamefont {Li}\ \emph {et~al.}(2016{\natexlab{a}})\citenamefont
  {Li}, \citenamefont {Neupert}, \citenamefont {Bernevig},\ and\ \citenamefont
  {Yazdani}}]{Li2016}%
  \BibitemOpen
  \bibfield  {author} {\bibinfo {author} {\bibfnamefont {J.}~\bibnamefont
  {Li}}, \bibinfo {author} {\bibfnamefont {T.}~\bibnamefont {Neupert}},
  \bibinfo {author} {\bibfnamefont {B.~A.}\ \bibnamefont {Bernevig}},\ and\
  \bibinfo {author} {\bibfnamefont {A.}~\bibnamefont {Yazdani}},\ }\bibfield
  {title} {\bibinfo {title} {{Manipulating Majorana zero modes on atomic rings
  with an external magnetic field}},\ }\href
  {https://doi.org/10.1038/ncomms10395} {\bibfield  {journal} {\bibinfo
  {journal} {Nature Communications}\ }\textbf {\bibinfo {volume} {7}},\
  \bibinfo {pages} {10395} (\bibinfo {year} {2016}{\natexlab{a}})}\BibitemShut
  {NoStop}%
\bibitem [{\citenamefont {Kreisel}\ \emph {et~al.}(2021)\citenamefont
  {Kreisel}, \citenamefont {Hyart},\ and\ \citenamefont {Rosenow}}]{Kreisel21}%
  \BibitemOpen
  \bibfield  {author} {\bibinfo {author} {\bibfnamefont {A.}~\bibnamefont
  {Kreisel}}, \bibinfo {author} {\bibfnamefont {T.}~\bibnamefont {Hyart}},\
  and\ \bibinfo {author} {\bibfnamefont {B.}~\bibnamefont {Rosenow}},\
  }\bibfield  {title} {\bibinfo {title} {{Tunable topological states hosted by
  unconventional superconductors with adatoms}},\ }\href
  {https://doi.org/10.1103/PhysRevResearch.3.033049} {\bibfield  {journal}
  {\bibinfo  {journal} {Phys. Rev. Research}\ }\textbf {\bibinfo {volume}
  {3}},\ \bibinfo {pages} {033049} (\bibinfo {year} {2021})}\BibitemShut
  {NoStop}%
\bibitem [{\citenamefont {Mishra}\ \emph {et~al.}(2021)\citenamefont {Mishra},
  \citenamefont {Simon}, \citenamefont {Hyart},\ and\ \citenamefont
  {Trif}}]{mishra2021yushibarusinov}%
  \BibitemOpen
  \bibfield  {author} {\bibinfo {author} {\bibfnamefont {A.}~\bibnamefont
  {Mishra}}, \bibinfo {author} {\bibfnamefont {P.}~\bibnamefont {Simon}},
  \bibinfo {author} {\bibfnamefont {T.}~\bibnamefont {Hyart}},\ and\ \bibinfo
  {author} {\bibfnamefont {M.}~\bibnamefont {Trif}},\ }\bibfield  {title}
  {\bibinfo {title} {{Yu-Shiba-Rusinov Qubit}},\ }\href
  {https://doi.org/10.1103/PRXQuantum.2.040347} {\bibfield  {journal} {\bibinfo
   {journal} {PRX Quantum}\ }\textbf {\bibinfo {volume} {2}},\ \bibinfo {pages}
  {040347} (\bibinfo {year} {2021})}\BibitemShut {NoStop}%
\bibitem [{\citenamefont {Nakosai}\ \emph {et~al.}(2013)\citenamefont
  {Nakosai}, \citenamefont {Tanaka},\ and\ \citenamefont
  {Nagaosa}}]{nakosai2013}%
  \BibitemOpen
  \bibfield  {author} {\bibinfo {author} {\bibfnamefont {S.}~\bibnamefont
  {Nakosai}}, \bibinfo {author} {\bibfnamefont {Y.}~\bibnamefont {Tanaka}},\
  and\ \bibinfo {author} {\bibfnamefont {N.}~\bibnamefont {Nagaosa}},\
  }\bibfield  {title} {\bibinfo {title} {{Two-dimensional p-wave
  superconducting states with magnetic moments on a conventional s-wave
  superconductor}},\ }\href
  {https://journals.aps.org/prb/abstract/10.1103/PhysRevB.88.180503} {\bibfield
   {journal} {\bibinfo  {journal} {Physical Review B}\ }\textbf {\bibinfo
  {volume} {88}},\ \bibinfo {pages} {180503} (\bibinfo {year}
  {2013})}\BibitemShut {NoStop}%
\bibitem [{\citenamefont {Kimme}\ \emph {et~al.}(2015)\citenamefont {Kimme},
  \citenamefont {Hyart},\ and\ \citenamefont {Rosenow}}]{Kimme2015}%
  \BibitemOpen
  \bibfield  {author} {\bibinfo {author} {\bibfnamefont {L.}~\bibnamefont
  {Kimme}}, \bibinfo {author} {\bibfnamefont {T.}~\bibnamefont {Hyart}},\ and\
  \bibinfo {author} {\bibfnamefont {B.}~\bibnamefont {Rosenow}},\ }\bibfield
  {title} {\bibinfo {title} {Symmetry-protected topological invariant and
  {M}ajorana impurity states in time-reversal-invariant superconductors},\
  }\href {https://doi.org/10.1103/PhysRevB.91.220501} {\bibfield  {journal}
  {\bibinfo  {journal} {Phys. Rev. B}\ }\textbf {\bibinfo {volume} {91}},\
  \bibinfo {pages} {220501(R)} (\bibinfo {year} {2015})}\BibitemShut {NoStop}%
\bibitem [{\citenamefont {Li}\ \emph {et~al.}(2016{\natexlab{b}})\citenamefont
  {Li}, \citenamefont {Neupert}, \citenamefont {Wang}, \citenamefont
  {MacDonald}, \citenamefont {Yazdani},\ and\ \citenamefont
  {Bernevig}}]{Bernevig16}%
  \BibitemOpen
  \bibfield  {author} {\bibinfo {author} {\bibfnamefont {J.}~\bibnamefont
  {Li}}, \bibinfo {author} {\bibfnamefont {T.}~\bibnamefont {Neupert}},
  \bibinfo {author} {\bibfnamefont {Z.}~\bibnamefont {Wang}}, \bibinfo {author}
  {\bibfnamefont {A.~H.}\ \bibnamefont {MacDonald}}, \bibinfo {author}
  {\bibfnamefont {A.}~\bibnamefont {Yazdani}},\ and\ \bibinfo {author}
  {\bibfnamefont {B.~A.}\ \bibnamefont {Bernevig}},\ }\bibfield  {title}
  {\bibinfo {title} {{Two-dimensional chiral topological superconductivity in
  Shiba lattices}},\ }\href {https://doi.org/10.1038/ncomms12297} {\bibfield
  {journal} {\bibinfo  {journal} {Nature Communications}\ }\textbf {\bibinfo
  {volume} {7}},\ \bibinfo {pages} {12297} (\bibinfo {year}
  {2016}{\natexlab{b}})}\BibitemShut {NoStop}%
\bibitem [{\citenamefont {Kimme}\ and\ \citenamefont {Hyart}(2016)}]{Kimme16}%
  \BibitemOpen
  \bibfield  {author} {\bibinfo {author} {\bibfnamefont {L.}~\bibnamefont
  {Kimme}}\ and\ \bibinfo {author} {\bibfnamefont {T.}~\bibnamefont {Hyart}},\
  }\bibfield  {title} {\bibinfo {title} {Existence of zero-energy impurity
  states in different classes of topological insulators and superconductors and
  their relation to topological phase transitions},\ }\href
  {https://doi.org/10.1103/PhysRevB.93.035134} {\bibfield  {journal} {\bibinfo
  {journal} {Phys. Rev. B}\ }\textbf {\bibinfo {volume} {93}},\ \bibinfo
  {pages} {035134} (\bibinfo {year} {2016})}\BibitemShut {NoStop}%
\bibitem [{\citenamefont {R\"ontynen}\ and\ \citenamefont
  {Ojanen}(2016)}]{Teemu16}%
  \BibitemOpen
  \bibfield  {author} {\bibinfo {author} {\bibfnamefont {J.}~\bibnamefont
  {R\"ontynen}}\ and\ \bibinfo {author} {\bibfnamefont {T.}~\bibnamefont
  {Ojanen}},\ }\bibfield  {title} {\bibinfo {title} {{Chern mosaic: Topology of
  chiral superconductivity on ferromagnetic adatom lattices}},\ }\href
  {https://doi.org/10.1103/PhysRevB.93.094521} {\bibfield  {journal} {\bibinfo
  {journal} {Phys. Rev. B}\ }\textbf {\bibinfo {volume} {93}},\ \bibinfo
  {pages} {094521} (\bibinfo {year} {2016})}\BibitemShut {NoStop}%
\bibitem [{\citenamefont {{Khosravian}}\ and\ \citenamefont
  {{Lado}}()}]{2022arXiv220211003K}%
  \BibitemOpen
  \bibfield  {author} {\bibinfo {author} {\bibfnamefont {M.}~\bibnamefont
  {{Khosravian}}}\ and\ \bibinfo {author} {\bibfnamefont {J.~L.}\ \bibnamefont
  {{Lado}}},\ }\bibfield  {title} {\bibinfo {title} {{{Impurity-induced
  excitations in twisted topological van der Waals superconductors}}},\
  }\href@noop {} {\ }\Eprint {https://arxiv.org/abs/2202.11003}
  {arXiv:2202.11003 [cond-mat.supr-con]} \BibitemShut {NoStop}%
\bibitem [{\citenamefont {Głodzik}\ and\ \citenamefont
  {Ojanen}(2020)}]{superYSR2020}%
  \BibitemOpen
  \bibfield  {author} {\bibinfo {author} {\bibfnamefont {S.}~\bibnamefont
  {Głodzik}}\ and\ \bibinfo {author} {\bibfnamefont {T.}~\bibnamefont
  {Ojanen}},\ }\bibfield  {title} {\bibinfo {title} {{Engineering nodal
  topological phases in Ising superconductors by magnetic superstructures}},\
  }\href {https://doi.org/10.1088/1367-2630/ab61d8} {\bibfield  {journal}
  {\bibinfo  {journal} {New Journal of Physics}\ }\textbf {\bibinfo {volume}
  {22}},\ \bibinfo {pages} {013022} (\bibinfo {year} {2020})}\BibitemShut
  {NoStop}%
\bibitem [{\citenamefont {Ptok}\ \emph {et~al.}(2017)\citenamefont {Ptok},
  \citenamefont {G\l{}odzik},\ and\ \citenamefont {Doma\ifmmode~\acute{n}\else
  \'{n}\fi{}ski}}]{PhysRevB.96.184425}%
  \BibitemOpen
  \bibfield  {author} {\bibinfo {author} {\bibfnamefont {A.}~\bibnamefont
  {Ptok}}, \bibinfo {author} {\bibfnamefont {S.}~\bibnamefont {G\l{}odzik}},\
  and\ \bibinfo {author} {\bibfnamefont {T.}~\bibnamefont
  {Doma\ifmmode~\acute{n}\else \'{n}\fi{}ski}},\ }\bibfield  {title} {\bibinfo
  {title} {{Yu-Shiba-Rusinov states of impurities in a triangular lattice of
  ${\mathrm{NbSe}}_{2}$ with spin-orbit coupling}},\ }\href
  {https://doi.org/10.1103/PhysRevB.96.184425} {\bibfield  {journal} {\bibinfo
  {journal} {Phys. Rev. B}\ }\textbf {\bibinfo {volume} {96}},\ \bibinfo
  {pages} {184425} (\bibinfo {year} {2017})}\BibitemShut {NoStop}%
\bibitem [{\citenamefont {Viyuela}\ \emph {et~al.}(2018)\citenamefont
  {Viyuela}, \citenamefont {Fu},\ and\ \citenamefont
  {Martin-Delgado}}]{PhysRevLett.120.017001}%
  \BibitemOpen
  \bibfield  {author} {\bibinfo {author} {\bibfnamefont {O.}~\bibnamefont
  {Viyuela}}, \bibinfo {author} {\bibfnamefont {L.}~\bibnamefont {Fu}},\ and\
  \bibinfo {author} {\bibfnamefont {M.~A.}\ \bibnamefont {Martin-Delgado}},\
  }\bibfield  {title} {\bibinfo {title} {{Chiral Topological Superconductors
  Enhanced by Long-Range Interactions}},\ }\href
  {https://doi.org/10.1103/PhysRevLett.120.017001} {\bibfield  {journal}
  {\bibinfo  {journal} {Phys. Rev. Lett.}\ }\textbf {\bibinfo {volume} {120}},\
  \bibinfo {pages} {017001} (\bibinfo {year} {2018})}\BibitemShut {NoStop}%
\bibitem [{\citenamefont {Thouless}\ \emph {et~al.}(1982)\citenamefont
  {Thouless}, \citenamefont {Kohmoto}, \citenamefont {Nightingale},\ and\
  \citenamefont {den Nijs}}]{TKNN}%
  \BibitemOpen
  \bibfield  {author} {\bibinfo {author} {\bibfnamefont {D.~J.}\ \bibnamefont
  {Thouless}}, \bibinfo {author} {\bibfnamefont {M.}~\bibnamefont {Kohmoto}},
  \bibinfo {author} {\bibfnamefont {M.~P.}\ \bibnamefont {Nightingale}},\ and\
  \bibinfo {author} {\bibfnamefont {M.}~\bibnamefont {den Nijs}},\ }\bibfield
  {title} {\bibinfo {title} {{Quantized Hall Conductance in a Two-Dimensional
  Periodic Potential}},\ }\href {https://doi.org/10.1103/PhysRevLett.49.405}
  {\bibfield  {journal} {\bibinfo  {journal} {Phys. Rev. Lett.}\ }\textbf
  {\bibinfo {volume} {49}},\ \bibinfo {pages} {405} (\bibinfo {year}
  {1982})}\BibitemShut {NoStop}%
\bibitem [{\citenamefont {Read}\ and\ \citenamefont {Green}(2000)}]{ReadGreen}%
  \BibitemOpen
  \bibfield  {author} {\bibinfo {author} {\bibfnamefont {N.}~\bibnamefont
  {Read}}\ and\ \bibinfo {author} {\bibfnamefont {D.}~\bibnamefont {Green}},\
  }\bibfield  {title} {\bibinfo {title} {{Paired states of fermions in two
  dimensions with breaking of parity and time-reversal symmetries and the
  fractional quantum Hall effect}},\ }\href
  {https://doi.org/10.1103/PhysRevB.61.10267} {\bibfield  {journal} {\bibinfo
  {journal} {Phys. Rev. B}\ }\textbf {\bibinfo {volume} {61}},\ \bibinfo
  {pages} {10267} (\bibinfo {year} {2000})}\BibitemShut {NoStop}%
\bibitem [{\citenamefont {Senthil}\ and\ \citenamefont
  {Fisher}(2000)}]{SenthilFisher}%
  \BibitemOpen
  \bibfield  {author} {\bibinfo {author} {\bibfnamefont {T.}~\bibnamefont
  {Senthil}}\ and\ \bibinfo {author} {\bibfnamefont {M.~P.~A.}\ \bibnamefont
  {Fisher}},\ }\bibfield  {title} {\bibinfo {title} {{Quasiparticle
  localization in superconductors with spin-orbit scattering}},\ }\href
  {https://doi.org/10.1103/PhysRevB.61.9690} {\bibfield  {journal} {\bibinfo
  {journal} {Phys. Rev. B}\ }\textbf {\bibinfo {volume} {61}},\ \bibinfo
  {pages} {9690} (\bibinfo {year} {2000})}\BibitemShut {NoStop}%
\bibitem [{\citenamefont {M{\'e}nard}\ \emph {et~al.}(2017)\citenamefont
  {M{\'e}nard}, \citenamefont {Guissart}, \citenamefont {Brun}, \citenamefont
  {Leriche}, \citenamefont {Trif}, \citenamefont {Debontridder}, \citenamefont
  {Demaille}, \citenamefont {Roditchev}, \citenamefont {Simon},\ and\
  \citenamefont {Cren}}]{Trif17}%
  \BibitemOpen
  \bibfield  {author} {\bibinfo {author} {\bibfnamefont {G.~C.}\ \bibnamefont
  {M{\'e}nard}}, \bibinfo {author} {\bibfnamefont {S.}~\bibnamefont
  {Guissart}}, \bibinfo {author} {\bibfnamefont {C.}~\bibnamefont {Brun}},
  \bibinfo {author} {\bibfnamefont {R.~T.}\ \bibnamefont {Leriche}}, \bibinfo
  {author} {\bibfnamefont {M.}~\bibnamefont {Trif}}, \bibinfo {author}
  {\bibfnamefont {F.}~\bibnamefont {Debontridder}}, \bibinfo {author}
  {\bibfnamefont {D.}~\bibnamefont {Demaille}}, \bibinfo {author}
  {\bibfnamefont {D.}~\bibnamefont {Roditchev}}, \bibinfo {author}
  {\bibfnamefont {P.}~\bibnamefont {Simon}},\ and\ \bibinfo {author}
  {\bibfnamefont {T.}~\bibnamefont {Cren}},\ }\bibfield  {title} {\bibinfo
  {title} {{Two-dimensional topological superconductivity in Pb/Co/Si(111)}},\
  }\href {https://doi.org/10.1038/s41467-017-02192-x} {\bibfield  {journal}
  {\bibinfo  {journal} {Nature Communications}\ }\textbf {\bibinfo {volume}
  {8}},\ \bibinfo {pages} {2040} (\bibinfo {year} {2017})}\BibitemShut
  {NoStop}%
\bibitem [{\citenamefont {Palacio-Morales}\ \emph {et~al.}(2019)\citenamefont
  {Palacio-Morales}, \citenamefont {Mascot}, \citenamefont {Cocklin},
  \citenamefont {Kim}, \citenamefont {Rachel}, \citenamefont {Morr},\ and\
  \citenamefont {Wiesendanger}}]{Wiesendager19}%
  \BibitemOpen
  \bibfield  {author} {\bibinfo {author} {\bibfnamefont {A.}~\bibnamefont
  {Palacio-Morales}}, \bibinfo {author} {\bibfnamefont {E.}~\bibnamefont
  {Mascot}}, \bibinfo {author} {\bibfnamefont {S.}~\bibnamefont {Cocklin}},
  \bibinfo {author} {\bibfnamefont {H.}~\bibnamefont {Kim}}, \bibinfo {author}
  {\bibfnamefont {S.}~\bibnamefont {Rachel}}, \bibinfo {author} {\bibfnamefont
  {D.~K.}\ \bibnamefont {Morr}},\ and\ \bibinfo {author} {\bibfnamefont
  {R.}~\bibnamefont {Wiesendanger}},\ }\bibfield  {title} {\bibinfo {title}
  {{Atomic-scale interface engineering of Majorana edge modes in a 2D
  magnet-superconductor hybrid system}},\ }\bibfield  {journal} {\bibinfo
  {journal} {Science Advances}\ }\textbf {\bibinfo {volume} {5}},\ \href
  {https://doi.org/10.1126/sciadv.aav6600} {10.1126/sciadv.aav6600} (\bibinfo
  {year} {2019})\BibitemShut {NoStop}%
\bibitem [{\citenamefont {Kezilebieke}\ \emph {et~al.}(2020)\citenamefont
  {Kezilebieke}, \citenamefont {Huda}, \citenamefont {Va{\v n}o}, \citenamefont
  {Aapro}, \citenamefont {Ganguli}, \citenamefont {Silveira}, \citenamefont
  {G{\l}odzik}, \citenamefont {Foster}, \citenamefont {Ojanen},\ and\
  \citenamefont {Liljeroth}}]{kezilebieke2020topological}%
  \BibitemOpen
  \bibfield  {author} {\bibinfo {author} {\bibfnamefont {S.}~\bibnamefont
  {Kezilebieke}}, \bibinfo {author} {\bibfnamefont {M.~N.}\ \bibnamefont
  {Huda}}, \bibinfo {author} {\bibfnamefont {V.}~\bibnamefont {Va{\v n}o}},
  \bibinfo {author} {\bibfnamefont {M.}~\bibnamefont {Aapro}}, \bibinfo
  {author} {\bibfnamefont {S.~C.}\ \bibnamefont {Ganguli}}, \bibinfo {author}
  {\bibfnamefont {O.~J.}\ \bibnamefont {Silveira}}, \bibinfo {author}
  {\bibfnamefont {S.}~\bibnamefont {G{\l}odzik}}, \bibinfo {author}
  {\bibfnamefont {A.~S.}\ \bibnamefont {Foster}}, \bibinfo {author}
  {\bibfnamefont {T.}~\bibnamefont {Ojanen}},\ and\ \bibinfo {author}
  {\bibfnamefont {P.}~\bibnamefont {Liljeroth}},\ }\bibfield  {title} {\bibinfo
  {title} {{Topological superconductivity in a van der Waals
  heterostructure}},\ }\href {https://doi.org/10.1038/s41586-020-2989-y}
  {\bibfield  {journal} {\bibinfo  {journal} {Nature}\ }\textbf {\bibinfo
  {volume} {588}},\ \bibinfo {pages} {424} (\bibinfo {year}
  {2020})}\BibitemShut {NoStop}%
\bibitem [{\citenamefont {Kezilebieke}\ \emph {et~al.}(2022)\citenamefont
  {Kezilebieke}, \citenamefont {Va{\v n}o}, \citenamefont {Huda}, \citenamefont
  {Aapro}, \citenamefont {Ganguli}, \citenamefont {Liljeroth},\ and\
  \citenamefont {Lado}}]{Kezilebieke22}%
  \BibitemOpen
  \bibfield  {author} {\bibinfo {author} {\bibfnamefont {S.}~\bibnamefont
  {Kezilebieke}}, \bibinfo {author} {\bibfnamefont {V.}~\bibnamefont {Va{\v
  n}o}}, \bibinfo {author} {\bibfnamefont {M.~N.}\ \bibnamefont {Huda}},
  \bibinfo {author} {\bibfnamefont {M.}~\bibnamefont {Aapro}}, \bibinfo
  {author} {\bibfnamefont {S.~C.}\ \bibnamefont {Ganguli}}, \bibinfo {author}
  {\bibfnamefont {P.}~\bibnamefont {Liljeroth}},\ and\ \bibinfo {author}
  {\bibfnamefont {J.~L.}\ \bibnamefont {Lado}},\ }\bibfield  {title} {\bibinfo
  {title} {{Moir\'e - Enabled Topological Superconductivity}},\ }\href
  {https://doi.org/10.1021/acs.nanolett.1c03856} {\bibfield  {journal}
  {\bibinfo  {journal} {Nano Letters}\ }\textbf {\bibinfo {volume} {22}},\
  \bibinfo {pages} {328} (\bibinfo {year} {2022})}\BibitemShut {NoStop}%
\bibitem [{\citenamefont {Carvalho}\ \emph {et~al.}(2018)\citenamefont
  {Carvalho}, \citenamefont {Garc\'{\i}a-Mart\'{\i}nez}, \citenamefont {Lado},\
  and\ \citenamefont {Fern\'andez-Rossier}}]{PhysRevB.97.115453}%
  \BibitemOpen
  \bibfield  {author} {\bibinfo {author} {\bibfnamefont {D.}~\bibnamefont
  {Carvalho}}, \bibinfo {author} {\bibfnamefont {N.~A.}\ \bibnamefont
  {Garc\'{\i}a-Mart\'{\i}nez}}, \bibinfo {author} {\bibfnamefont {J.~L.}\
  \bibnamefont {Lado}},\ and\ \bibinfo {author} {\bibfnamefont
  {J.}~\bibnamefont {Fern\'andez-Rossier}},\ }\bibfield  {title} {\bibinfo
  {title} {Real-space mapping of topological invariants using artificial neural
  networks},\ }\href {https://doi.org/10.1103/PhysRevB.97.115453} {\bibfield
  {journal} {\bibinfo  {journal} {Phys. Rev. B}\ }\textbf {\bibinfo {volume}
  {97}},\ \bibinfo {pages} {115453} (\bibinfo {year} {2018})}\BibitemShut
  {NoStop}%
\bibitem [{\citenamefont {Rodriguez-Nieva}\ and\ \citenamefont
  {Scheurer}(2019)}]{MLphase2019}%
  \BibitemOpen
  \bibfield  {author} {\bibinfo {author} {\bibfnamefont {J.~F.}\ \bibnamefont
  {Rodriguez-Nieva}}\ and\ \bibinfo {author} {\bibfnamefont {M.~S.}\
  \bibnamefont {Scheurer}},\ }\bibfield  {title} {\bibinfo {title}
  {{Identifying topological order through unsupervised machine learning}},\
  }\href {https://doi.org/10.1038/s41567-019-0512-x} {\bibfield  {journal}
  {\bibinfo  {journal} {Nature Physics}\ }\textbf {\bibinfo {volume} {15}},\
  \bibinfo {pages} {790–795} (\bibinfo {year} {2019})}\BibitemShut {NoStop}%
\bibitem [{\citenamefont {Holanda}\ and\ \citenamefont
  {Griffith}(2020)}]{PhysRevB.102.054107}%
  \BibitemOpen
  \bibfield  {author} {\bibinfo {author} {\bibfnamefont {N.~L.}\ \bibnamefont
  {Holanda}}\ and\ \bibinfo {author} {\bibfnamefont {M.~A.~R.}\ \bibnamefont
  {Griffith}},\ }\bibfield  {title} {\bibinfo {title} {{Machine learning
  topological phases in real space}},\ }\href
  {https://doi.org/10.1103/PhysRevB.102.054107} {\bibfield  {journal} {\bibinfo
   {journal} {Phys. Rev. B}\ }\textbf {\bibinfo {volume} {102}},\ \bibinfo
  {pages} {054107} (\bibinfo {year} {2020})}\BibitemShut {NoStop}%
\bibitem [{\citenamefont {Lustig}\ \emph {et~al.}(2020)\citenamefont {Lustig},
  \citenamefont {Yair}, \citenamefont {Talmon},\ and\ \citenamefont
  {Segev}}]{PhysRevLett.125.127401}%
  \BibitemOpen
  \bibfield  {author} {\bibinfo {author} {\bibfnamefont {E.}~\bibnamefont
  {Lustig}}, \bibinfo {author} {\bibfnamefont {O.}~\bibnamefont {Yair}},
  \bibinfo {author} {\bibfnamefont {R.}~\bibnamefont {Talmon}},\ and\ \bibinfo
  {author} {\bibfnamefont {M.}~\bibnamefont {Segev}},\ }\bibfield  {title}
  {\bibinfo {title} {{Identifying Topological Phase Transitions in Experiments
  Using Manifold Learning}},\ }\href
  {https://doi.org/10.1103/PhysRevLett.125.127401} {\bibfield  {journal}
  {\bibinfo  {journal} {Phys. Rev. Lett.}\ }\textbf {\bibinfo {volume} {125}},\
  \bibinfo {pages} {127401} (\bibinfo {year} {2020})}\BibitemShut {NoStop}%
\bibitem [{\citenamefont {Baireuther}\ \emph {et~al.}()\citenamefont
  {Baireuther}, \citenamefont {Płodzień}, \citenamefont {Ojanen},
  \citenamefont {Tworzydło},\ and\ \citenamefont {Hyart}}]{baireuther2021}%
  \BibitemOpen
  \bibfield  {author} {\bibinfo {author} {\bibfnamefont {P.}~\bibnamefont
  {Baireuther}}, \bibinfo {author} {\bibfnamefont {M.}~\bibnamefont
  {Płodzień}}, \bibinfo {author} {\bibfnamefont {T.}~\bibnamefont {Ojanen}},
  \bibinfo {author} {\bibfnamefont {J.}~\bibnamefont {Tworzydło}},\ and\
  \bibinfo {author} {\bibfnamefont {T.}~\bibnamefont {Hyart}},\ }\href@noop {}
  {\bibinfo {title} {{Identifying Chern numbers of superconductors from local
  measurements}}},\ \Eprint {https://arxiv.org/abs/2112.06777}
  {arXiv:2112.06777 [cond-mat.mes-hall]} \BibitemShut {NoStop}%
\bibitem [{\citenamefont {P{\"o}yh{\"o}nen}\ \emph {et~al.}(2018)\citenamefont
  {P{\"o}yh{\"o}nen}, \citenamefont {Sahlberg}, \citenamefont {Weststr{\"o}m},\
  and\ \citenamefont {Ojanen}}]{Kim2018}%
  \BibitemOpen
  \bibfield  {author} {\bibinfo {author} {\bibfnamefont {K.}~\bibnamefont
  {P{\"o}yh{\"o}nen}}, \bibinfo {author} {\bibfnamefont {I.}~\bibnamefont
  {Sahlberg}}, \bibinfo {author} {\bibfnamefont {A.}~\bibnamefont
  {Weststr{\"o}m}},\ and\ \bibinfo {author} {\bibfnamefont {T.}~\bibnamefont
  {Ojanen}},\ }\bibfield  {title} {\bibinfo {title} {{Amorphous topological
  superconductivity in a Shiba glass}},\ }\href
  {https://doi.org/10.1038/s41467-018-04532-x} {\bibfield  {journal} {\bibinfo
  {journal} {Nature Communications}\ }\textbf {\bibinfo {volume} {9}},\
  \bibinfo {pages} {2103} (\bibinfo {year} {2018})}\BibitemShut {NoStop}%
\bibitem [{pyq()}]{pyqula}%
  \BibitemOpen
  \href@noop {} {\bibinfo  {journal} {\mbox{pyqula Library}
  https://github.com/joselado/pyqula}\ }\BibitemShut {NoStop}%
\bibitem [{\citenamefont {Fukui}\ \emph {et~al.}(2005)\citenamefont {Fukui},
  \citenamefont {Hatsugai},\ and\ \citenamefont {Suzuki}}]{Fukui05}%
  \BibitemOpen
\bibfield  {journal} {  }\bibfield  {author} {\bibinfo {author} {\bibfnamefont
  {T.}~\bibnamefont {Fukui}}, \bibinfo {author} {\bibfnamefont
  {Y.}~\bibnamefont {Hatsugai}},\ and\ \bibinfo {author} {\bibfnamefont
  {H.}~\bibnamefont {Suzuki}},\ }\bibfield  {title} {\bibinfo {title} {{Chern
  Numbers in Discretized Brillouin Zone: Efficient Method of Computing (Spin)
  Hall Conductances}},\ }\href {https://doi.org/10.1143/JPSJ.74.1674}
  {\bibfield  {journal} {\bibinfo  {journal} {Journal of the Physical Society
  of Japan}\ }\textbf {\bibinfo {volume} {74}},\ \bibinfo {pages} {1674}
  (\bibinfo {year} {2005})}\BibitemShut {NoStop}%
\bibitem [{\citenamefont {Lee}(2018)}]{FL2018}%
  \BibitemOpen
  \bibfield  {author} {\bibinfo {author} {\bibfnamefont {S.-S.}\ \bibnamefont
  {Lee}},\ }\bibfield  {title} {\bibinfo {title} {{Recent Developments in
  Non-Fermi Liquid Theory}},\ }\href
  {https://doi.org/10.1146/annurev-conmatphys-031016-025531} {\bibfield
  {journal} {\bibinfo  {journal} {Annual Review of Condensed Matter Physics}\
  }\textbf {\bibinfo {volume} {9}},\ \bibinfo {pages} {227–244} (\bibinfo
  {year} {2018})}\BibitemShut {NoStop}%
\bibitem [{\citenamefont {Parisen~Toldin}\ \emph {et~al.}(2015)\citenamefont
  {Parisen~Toldin}, \citenamefont {Hohenadler}, \citenamefont {Assaad},\ and\
  \citenamefont {Herbut}}]{Hubbard2015}%
  \BibitemOpen
  \bibfield  {author} {\bibinfo {author} {\bibfnamefont {F.}~\bibnamefont
  {Parisen~Toldin}}, \bibinfo {author} {\bibfnamefont {M.}~\bibnamefont
  {Hohenadler}}, \bibinfo {author} {\bibfnamefont {F.~F.}\ \bibnamefont
  {Assaad}},\ and\ \bibinfo {author} {\bibfnamefont {I.~F.}\ \bibnamefont
  {Herbut}},\ }\bibfield  {title} {\bibinfo {title} {{Fermionic quantum
  criticality in honeycomb and $\ensuremath{\pi}$-flux Hubbard models:
  Finite-size scaling of renormalization-group-invariant observables from
  quantum Monte Carlo}},\ }\href {https://doi.org/10.1103/PhysRevB.91.165108}
  {\bibfield  {journal} {\bibinfo  {journal} {Phys. Rev. B}\ }\textbf {\bibinfo
  {volume} {91}},\ \bibinfo {pages} {165108} (\bibinfo {year}
  {2015})}\BibitemShut {NoStop}%
\bibitem [{\citenamefont {Moca}\ \emph {et~al.}(2021)\citenamefont {Moca},
  \citenamefont {Weymann}, \citenamefont {Werner},\ and\ \citenamefont
  {Zar\'and}}]{PhysRevLett.127.186804}%
  \BibitemOpen
  \bibfield  {author} {\bibinfo {author} {\bibfnamefont {C.}~\bibnamefont
  {Moca}}, \bibinfo {author} {\bibfnamefont {I.}~\bibnamefont {Weymann}},
  \bibinfo {author} {\bibfnamefont {M.~A.}\ \bibnamefont {Werner}},\ and\
  \bibinfo {author} {\bibfnamefont {G.}~\bibnamefont {Zar\'and}},\ }\bibfield
  {title} {\bibinfo {title} {{Kondo Cloud in a Superconductor}},\ }\href
  {https://doi.org/10.1103/PhysRevLett.127.186804} {\bibfield  {journal}
  {\bibinfo  {journal} {Phys. Rev. Lett.}\ }\textbf {\bibinfo {volume} {127}},\
  \bibinfo {pages} {186804} (\bibinfo {year} {2021})}\BibitemShut {NoStop}%
\bibitem [{\citenamefont {Abouelkomsan}\ \emph {et~al.}(2020)\citenamefont
  {Abouelkomsan}, \citenamefont {Liu},\ and\ \citenamefont
  {Bergholtz}}]{PhysRevLett.124.106803}%
  \BibitemOpen
  \bibfield  {author} {\bibinfo {author} {\bibfnamefont {A.}~\bibnamefont
  {Abouelkomsan}}, \bibinfo {author} {\bibfnamefont {Z.}~\bibnamefont {Liu}},\
  and\ \bibinfo {author} {\bibfnamefont {E.~J.}\ \bibnamefont {Bergholtz}},\
  }\bibfield  {title} {\bibinfo {title} {{Particle-Hole Duality, Emergent Fermi
  Liquids, and Fractional Chern Insulators in Moir\'e Flatbands}},\ }\href
  {https://doi.org/10.1103/PhysRevLett.124.106803} {\bibfield  {journal}
  {\bibinfo  {journal} {Phys. Rev. Lett.}\ }\textbf {\bibinfo {volume} {124}},\
  \bibinfo {pages} {106803} (\bibinfo {year} {2020})}\BibitemShut {NoStop}%
\bibitem [{\citenamefont {Repellin}\ and\ \citenamefont
  {Senthil}(2020)}]{PhysRevResearch.2.023238}%
  \BibitemOpen
  \bibfield  {author} {\bibinfo {author} {\bibfnamefont {C.}~\bibnamefont
  {Repellin}}\ and\ \bibinfo {author} {\bibfnamefont {T.}~\bibnamefont
  {Senthil}},\ }\bibfield  {title} {\bibinfo {title} {{Chern bands of twisted
  bilayer graphene: Fractional Chern insulators and spin phase transition}},\
  }\href {https://doi.org/10.1103/PhysRevResearch.2.023238} {\bibfield
  {journal} {\bibinfo  {journal} {Phys. Rev. Research}\ }\textbf {\bibinfo
  {volume} {2}},\ \bibinfo {pages} {023238} (\bibinfo {year}
  {2020})}\BibitemShut {NoStop}%
\bibitem [{\citenamefont {Orús}(2014)}]{Orus2014}%
  \BibitemOpen
  \bibfield  {author} {\bibinfo {author} {\bibfnamefont {R.}~\bibnamefont
  {Orús}},\ }\bibfield  {title} {\bibinfo {title} {{A practical introduction
  to tensor networks: Matrix product states and projected entangled pair
  states}},\ }\href {https://doi.org/10.1016/j.aop.2014.06.013} {\bibfield
  {journal} {\bibinfo  {journal} {Annals of Physics}\ }\textbf {\bibinfo
  {volume} {349}},\ \bibinfo {pages} {117–158} (\bibinfo {year}
  {2014})}\BibitemShut {NoStop}%
\bibitem [{\citenamefont {Carleo}\ and\ \citenamefont
  {Troyer}(2017)}]{Carleo2017}%
  \BibitemOpen
  \bibfield  {author} {\bibinfo {author} {\bibfnamefont {G.}~\bibnamefont
  {Carleo}}\ and\ \bibinfo {author} {\bibfnamefont {M.}~\bibnamefont
  {Troyer}},\ }\bibfield  {title} {\bibinfo {title} {{Solving the quantum
  many-body problem with artificial neural networks}},\ }\href
  {https://doi.org/10.1126/science.aag2302} {\bibfield  {journal} {\bibinfo
  {journal} {Science}\ }\textbf {\bibinfo {volume} {355}},\ \bibinfo {pages}
  {602–606} (\bibinfo {year} {2017})}\BibitemShut {NoStop}%
\bibitem [{\citenamefont {Georges}\ \emph {et~al.}(1996)\citenamefont
  {Georges}, \citenamefont {Kotliar}, \citenamefont {Krauth},\ and\
  \citenamefont {Rozenberg}}]{RevModPhys.68.13}%
  \BibitemOpen
  \bibfield  {author} {\bibinfo {author} {\bibfnamefont {A.}~\bibnamefont
  {Georges}}, \bibinfo {author} {\bibfnamefont {G.}~\bibnamefont {Kotliar}},
  \bibinfo {author} {\bibfnamefont {W.}~\bibnamefont {Krauth}},\ and\ \bibinfo
  {author} {\bibfnamefont {M.~J.}\ \bibnamefont {Rozenberg}},\ }\bibfield
  {title} {\bibinfo {title} {{Dynamical mean-field theory of strongly
  correlated fermion systems and the limit of infinite dimensions}},\ }\href
  {https://doi.org/10.1103/RevModPhys.68.13} {\bibfield  {journal} {\bibinfo
  {journal} {Rev. Mod. Phys.}\ }\textbf {\bibinfo {volume} {68}},\ \bibinfo
  {pages} {13} (\bibinfo {year} {1996})}\BibitemShut {NoStop}%
\bibitem [{\citenamefont {Onida}\ \emph {et~al.}(2002)\citenamefont {Onida},
  \citenamefont {Reining},\ and\ \citenamefont {Rubio}}]{RevModPhys.74.601}%
  \BibitemOpen
  \bibfield  {author} {\bibinfo {author} {\bibfnamefont {G.}~\bibnamefont
  {Onida}}, \bibinfo {author} {\bibfnamefont {L.}~\bibnamefont {Reining}},\
  and\ \bibinfo {author} {\bibfnamefont {A.}~\bibnamefont {Rubio}},\ }\bibfield
   {title} {\bibinfo {title} {{Electronic excitations: density-functional
  versus many-body Green's-function approaches}},\ }\href
  {https://doi.org/10.1103/RevModPhys.74.601} {\bibfield  {journal} {\bibinfo
  {journal} {Rev. Mod. Phys.}\ }\textbf {\bibinfo {volume} {74}},\ \bibinfo
  {pages} {601} (\bibinfo {year} {2002})}\BibitemShut {NoStop}%
\bibitem [{\citenamefont {Stoudenmire}\ \emph {et~al.}(2011)\citenamefont
  {Stoudenmire}, \citenamefont {Alicea}, \citenamefont {Starykh},\ and\
  \citenamefont {Fisher}}]{Stoudenmire}%
  \BibitemOpen
  \bibfield  {author} {\bibinfo {author} {\bibfnamefont {E.~M.}\ \bibnamefont
  {Stoudenmire}}, \bibinfo {author} {\bibfnamefont {J.}~\bibnamefont {Alicea}},
  \bibinfo {author} {\bibfnamefont {O.~A.}\ \bibnamefont {Starykh}},\ and\
  \bibinfo {author} {\bibfnamefont {M.~P.}\ \bibnamefont {Fisher}},\ }\bibfield
   {title} {\bibinfo {title} {{Interaction effects in topological
  superconducting wires supporting Majorana fermions}},\ }\href
  {https://doi.org/10.1103/PhysRevB.84.014503} {\bibfield  {journal} {\bibinfo
  {journal} {Phys. Rev. B}\ }\textbf {\bibinfo {volume} {84}},\ \bibinfo
  {pages} {014503} (\bibinfo {year} {2011})}\BibitemShut {NoStop}%
\end{thebibliography}%

\end{document}